\documentclass[a4paper, 11pt]{article}

\usepackage{jheppub}
\usepackage{abb}

\usepackage{amsmath, amsfonts}
\usepackage[normalem]{ulem}
\usepackage{bm}
\usepackage{physics}

\usepackage{comment}
\usepackage{color}

\usepackage[framemethod=default]{mdframed}
\newmdenv[skipabove=7pt,
skipbelow=7pt,
rightline=false,
leftline=false,
topline=false,
bottomline=false,
backgroundcolor=gray!10,
linecolor=gray,
innerleftmargin=5pt,
innerrightmargin=5pt,
innertopmargin=5pt,
innerbottommargin=5pt,
leftmargin=0cm,
rightmargin=0cm,
linewidth=4pt]{eBox}

\makeatletter
\@addtoreset{equation}{section}

\makeatother

\allowdisplaybreaks[4]
\usepackage{cancel}
\usepackage{bbold}
\usepackage{empheq}
\usepackage{framed}
\usepackage[most]{tcolorbox}
\usepackage{xcolor}
\colorlet{shadecolor}{orange!15}
\usepackage{braket}
\usepackage{caption}
\usepackage{subcaption}
\usepackage{tikz}
\definecolor{pyblue}{RGB}{31, 119, 180}
\definecolor{pyorange}{RGB}{255, 127, 14}
\definecolor{pygreen}{RGB}{44, 160, 44}
\definecolor{pyred}{RGB}{214, 39, 40}
\usetikzlibrary{intersections, calc, arrows.meta}
\usetikzlibrary{patterns}

\usepackage{array}
\usepackage{longtable}

\usepackage{adjustbox}
\usepackage{tikz-feynman}
\definecolor{rossocorsa}{rgb}{0.83, 0.0, 0.0}

\definecolor{gbcolor2}{rgb}{.9,.2,.6}
\definecolor{gbcolor3}{rgb}{.3,.2,.6}

\definecolor{verdechiaro}{rgb}{0.6,1,0.6}
\definecolor{giallochiaro}{rgb}{1,1,0.6}
\definecolor{bluscuro}{rgb}{0.15, 0.2, 0.9}
\definecolor{verdes}{rgb}{0.1, 0.5, 0.1}
\definecolor{tangerineyellow}{rgb}{1.0, 0.8, 0.0}
\definecolor{smokyblack}{rgb}{0.06, 0.05, 0.03}
\definecolor{americanrose}{rgb}{1.0, 0.01, 0.24}
\definecolor{cobalt}{rgb}{0.0, 0.28, 0.67}
\definecolor{brandeisblue}{rgb}{0.0, 0.44, 1.0}
\definecolor{mycolor}{rgb}{0.0, 0.0, 0.5}
\definecolor{oxfordblue}{rgb}{0.0, 0.13, 0.28}
\definecolor{azure}{rgb}{0.0, 0.5, 1.0}
\definecolor{turquoiseblue}{rgb}{0.0, 1.0, 0.94}
\definecolor{venetianred}{rgb}{0.78, 0.03, 0.08}
\newtcolorbox{mynamedbox1}[1]{colback=venetianred!5!white,colframe=venetianred!80!black,title=#1}
\usepackage{amssymb}
\usepackage{pifont}

\newcommand\mpl{M_{\rm Pl}}

\usepackage[symbol]{footmisc}
\newcommand{\ctext}[1]{\raise0.2ex\hbox{\textcircled{\scriptsize{#1}}}}

\preprint{CERN-TH-2026-038}
\title{
\boldmath Scale-Dependent Loop Corrections to the Inflationary Power Spectrum}
\author[a]{Matteo Braglia}
\author[b]{, Sebastián Céspedes}
\author[c]{and  Lucas Pinol}

\affiliation[a]{Theoretical Physics Department, CERN, 1211 Geneva 23, Switzerland}
\affiliation[b]{Abdus Salam Centre for Theoretical Physics, Imperial College London, London, SW7 2AZ, UK}
\affiliation[c]{ Laboratoire de Physique de l’École Normale Supérieure, ENS, CNRS, Université PSL,\\ Sorbonne Université, Université Paris Cité, F-75005, Paris, France}

\emailAdd{matteo.braglia@cern.ch}
\emailAdd{s.cespedes-castillo@imperial.ac.uk}
\emailAdd{lucas.pinol@phys.ens.fr}

\abstract{
Loop corrections to primordial correlation functions are unavoidable due to the non-linear nature of gravity. Previous works have established a robust framework for computing the renormalised one-loop power spectra of scalar and tensor modes, but primarily in (near) de Sitter backgrounds. In this work, we develop a consistent renormalisation procedure applicable to inflationary backgrounds that strongly break de Sitter symmetries and generate scale-dependent features in the primordial spectra. Our analysis is performed within the Effective Field Theory (EFT) of inflationary fluctuations, allowing for arbitrary time dependence of the Wilson coefficients.
We show that both ultraviolet divergences and tadpoles of the theory, despite their strong time and scale dependence, can be cancelled by a finite set of local counter-terms compatible with the EFT symmetries. Importantly, this result only relies on the existence of an initial phase of adiabatic evolution continuously related to the Bunch-Davies vacuum and holds independently of the precise time dependence of the background and of the free-field mode functions. 
We then study two concrete realisations, corresponding to resonant and sharp features. In both cases, all calculations are carried out exactly in the limit of small feature amplitude.
We analyse perturbativity and provide the first explicit demonstration that the renormalised one-loop power spectrum generated by a localised feature along the inflationary trajectory vanishes both at large and small scales. 
Our scale-dependent renormalisation framework implies that models of primordial features used to fit CMB residuals are consistent with perturbativity bounds, and opens the door to systematic studies of loop corrections in more complicated scenarios relevant for scalar-induced gravitational waves and primordial black holes.}

\begin{document}

\maketitle
\renewcommand{\thefootnote}{\arabic{footnote}}
\setcounter{footnote}{0}
\pagebreak

\section{Introduction}
Obtaining precision predictions for inflationary observables is one of the central goals of modern cosmology. Beyond their phenomenological relevance, such predictions probe the consistency of quantum field theory in time-dependent backgrounds and the validity of perturbation theory in an expanding universe. While tree-level predictions for the scalar power spectrum and higher-point correlators are by now well understood, the role and interpretation of loop corrections remain subtle.

Pioneering works by Weinberg initiated the systematic study of loop effects in inflationary spacetimes and clarified that quantum corrections in an expanding background can exhibit features that differ qualitatively from their flat-spacetime counterparts, including enhanced sensitivity to infrared modes and potentially non-trivial time dependence \cite{Weinberg:2005vy,Weinberg:2006ac}. These early analyses raised foundational questions regarding the late-time behaviour of cosmological correlators and the consistency of perturbation theory in de Sitter and quasi–de Sitter spacetimes.

Subsequent works refined this picture and demonstrated that, in single-clock slow-roll inflation, late-time correlators of light fields approach constant values to all orders in perturbation theory. In particular, it was shown that no secular time dependence survives at late times once all contributions are consistently included \cite{Senatore:2012ya,Assassi:2012et}. This result--explicitly proved at 1-loop level in~\cite{Braglia:2025cee,Braglia:2025qrb}--indicates that, despite the presence of interactions and loop corrections, inflationary correlators freeze outside the horizon and do not evolve at late times.

The absence of secular late-time evolution suggests that quantum corrections do not accumulate as modes freeze, and that inflationary dynamics is effectively adiabatic in the slow-roll regime. However, this observation alone does not determine whether loop corrections are physically meaningful or observable. Even if correlators are time independent at late times, it remains unclear whether loop corrections can be operationally distinguished from tree-level results once parameters are fixed by renormalisation conditions. In particular, the absence of secular growth does not, by itself, imply that loop effects can always be absorbed into a redefinition of background quantities.

This question admits a natural interpretation within the effective field theory (EFT) of inflationary perturbations, where the dynamics of scalar perturbations is organised in terms of operators consistent with the non-linearly realised time diffeomorphisms \cite{Creminelli:2006xe,Cheung:2007st}. In this framework, the breaking of time-translation invariance is set by the background evolution and is therefore approximately adiabatic in slow-roll inflation. Loop corrections renormalise the coefficients of operators already present in the EFT, and the absence of late-time evolution of correlators suggests that quantum effects do not generate explicit running or new time dependence in observables evaluated at late times. Renormalisability is understood here in the EFT sense, namely that loop divergences can be absorbed into a finite set of operators consistent with the symmetries and power counting of the effective theory.

A particularly transparent way of implementing this EFT is provided by the decoupling limit, in which the mixing between scalar fluctuations and metric perturbations can be neglected. In the EFT of inflation the dynamics is captured by the Goldstone mode $\pi$ associated with the spontaneous breaking of time translations \cite{Cheung:2007st}. In this limit, the same EFT structure is expressed in terms of a Lagrangian for $\pi$ with explicitly time-dependent couplings multiplying operators such as $\dot\pi^2$, $(\partial_i\pi)^2$, and higher-order interactions including $\dot\pi^3$ and $\dot\pi(\partial_i\pi)^2$. Loop corrections arise from these interactions and are computed using the in-in formalism. The question of interest can then be phrased sharply: do quantum corrections to the $\pi$ two-point function generate effects that cannot be absorbed into a renormalisation of EFT coefficients already present at tree level? Do they merely reshuffle parameters without leading to new physical time or scale dependence in late-time observables?
Does one need to study higher-point correlation functions, as recently suggested in~\cite{Ballesteros:2025nhz}?
Our emphasis in this work is not only on the numerical size of loop corrections, but on their structural role within the EFT.

Addressing this question requires explicit control over loop computations in inflationary backgrounds and within the EFT of inflation, particularly in the presence of time-dependent interactions. A substantial body of work has developed complementary techniques to organise in-in perturbation theory and renormalisation in quasi–de Sitter spacetimes, including systematic analyses of ultraviolet (UV) divergences, counter-terms, and the late-time behaviour of cosmological correlators~\cite{Senatore:2009cf,Marolf:2010zp,Senatore:2012nq,Premkumar:2021mlz,Xianyu:2022jwk,Qin:2023bjk,Qin:2024gtr,Cespedes:2025dnq,Braglia:2025qrb,Braglia:2025cee,Cespedes:2025ple,Ballesteros:2025nhz,Pimentel:2026kqc,Ema:2026dop,Farren:2026hao}. These developments provide the conceptual and technical foundations for treating loop corrections consistently in inflationary settings, including scenarios with explicitly time-dependent couplings and scale-dependent backgrounds.

Building on these recent frameworks, 
as well as
a technique introduced 
in~\cite{Ballesteros:2024cef}
to compute in-in integrals in dimensional regularisation,~\cite{Braglia:2025cee,Braglia:2025qrb} provided a definitive answer to this question in the context of slow-roll inflation. 
There it was explicitly shown that both the renormalised scalar--and scalar-induced tensor--power spectrum at 1-loop are indistinguishable from the tree-level result. The case of the scalar-induced tensor-power spectrum was previously solved in Ref.~\cite{Ballesteros:2024cef}, whose results agree with those of~\cite{Braglia:2025cee}.  
In this sense, slow-roll inflation is radiatively stable in a strong operational sense, as quantum corrections do not lead to observable modifications of the late-time power spectrum. 
In this paper, we extend this line of investigation to situations in which the EFT of inflation contains an additional physical time scale controlling the time dependence of the interactions. Such scenarios, often referred to as {\em primordial features}, go beyond the adiabatic slow-roll regime and involve a more explicit breaking of time-translation invariance. They are a natural prediction of many UV-complete inflationary models~\cite{Chen:2010xka,Achucarro:2022qrl} and, although they have not been detected with statistical significance to date, they improve the fit to CMB data compared to slow-roll scenarios~\cite{Planck:2018jri}. We focus on two representative examples; resonant and sharp features, which correspond to small oscillatory and localised corrections to the inflationary background, respectively\footnote{Another common type of primordial feature is the so-called classical primordial standard clock, generated when a massive field classically oscillates during inflation~\cite{Chen:2014cwa,Braglia:2021ckn}. However, these models are inherently multifield, and modeling such a time dependence of the background requires going beyond the single-field EFT (see, e.g.,~\cite{Pinol:2024arz}), which is outside the scope of this paper.}.
Although both cases introduce time-dependent couplings within the EFT, they do so in qualitatively different ways, and this difference can be traced to symmetry considerations rather than to the detailed form of the background.

We first show that, independently of the detailed background evolution, the theory remains renormalisable in the EFT sense. In particular, loop divergences can always be absorbed into a finite set of counter-terms whose structure is fixed by the symmetries and power counting of the effective action. An important technical aspect of this analysis is that operators which are redundant at tree level must be retained in the presence of time-dependent couplings in order to achieve a consistent renormalisation. This result establishes the closure of the EFT under loops even in the presence of explicit time dependence.

Within this general framework, we find that in the resonant case loop corrections to the scalar two-point function are indistinguishable from the tree-level result. The underlying reason is that the oscillatory background preserves a remnant discrete shift symmetry, which constrains the effective action and prevents loop effects from generating observable corrections beyond a renormalisation of EFT parameters.

This conclusion does not extend to the case of sharp features. In that case, the explicit and localised breaking of time-translation invariance is not accompanied by a residual symmetry, and loop corrections are therefore less constrained. As a result, loop effects can induce a phase shift in the features of the two-point function that cannot be absorbed into a redefinition of tree-level parameters. Although such effects are suppressed by the cut-off of the EFT, they correspond to a genuine loop-induced modification of the power spectrum. In contrast to earlier concerns about secular growth in inflationary loops, we find that in our scale-dependent scenarios secular divergences are regulated by the feature itself and do not control the late-time behaviour. Moreover, once renormalised, the one-loop power spectrum vanishes both at scales much smaller and much larger than the one of the feature.

The paper is organised as follows. In Section~\ref{sec2} we review the effective field theory of inflation, introduce the counter-terms relevant to our analysis, and specify the region of parameter space that we will consider for practical purposes.
Section~\ref{sec3} is devoted to the general structure of loop corrections in time-dependent inflationary backgrounds. We present the bare one-loop computation, highlight the main differences with respect to the scale-invariant case, and discuss the renormalisation procedure, including the role of tadpole contributions required for a consistent calculation at the order considered. In Section~\ref{sec4} we apply this framework to two explicit models. We first present the corresponding tree-level predictions for the scalar power spectrum, fixing conventions and approximations relevant for the loop analysis. We then compute the fully renormalised one-loop power spectrum, including the combined effects of bare loops, quadratic counter-terms, and tadpole-induced contributions. Finally, we compare the resulting scale dependence with the tree-level spectrum and discuss its interpretation, observability, and the regime of validity of the EFT, including perturbativity bounds on the feature parameters. Further technical details and explicit expressions are presented in several appendices.

\section{Loop Corrections in the Effective Field Theory of Inflation}
\label{sec2}
\subsection{Effective Field Theory of Inflation}
A powerful framework to parametrise the interactions of perturbations during inflation is the effective field theory (EFT) of inflation, formulated in terms of the Goldstone mode associated with spontaneously broken time diffeomorphisms \cite{Creminelli:2006xe, Cheung:2007st}. The key observation is that an inflationary background selects a preferred time slicing: the homogeneous solution for the inflaton $\phi_0(t)$, or equivalently the Hubble rate $H(t)$, is explicitly time-dependent. As a result, time translations are spontaneously broken by the background, while spatial diffeomorphisms remain unbroken.

The dynamics of fluctuations can therefore be organised according to this symmetry-breaking pattern. The EFT approach constructs the most general action compatible with the unbroken symmetries. Once time diffeomorphisms are broken, the action need only remain invariant under spatial diffeomorphisms, with time diffeomorphisms realised non-linearly. Assuming locality, this symmetry structure fixes the allowed operator basis and provides a systematic expansion in derivatives and powers of the fields.

The construction proceeds in unitary gauge, where fluctuations of the clock field are set to zero and all degrees of freedom are encoded in the metric. In this gauge the action is invariant under time-dependent spatial diffeomorphisms and contains arbitrary functions of time multiplying operators such as $g^{00}$, the extrinsic curvature $K_{ij}$, and higher-derivative combinations. The explicit time dependence of these coefficients reflects the spontaneous breaking of time translations by the background.

Time diffeomorphism invariance is then restored using the Stückelberg trick by performing the replacement
$t \to t + \pi(t,\bm{x})$.
Under an infinitesimal time translation $t \to t + \xi^0$, the Goldstone field transforms non-linearly as $\pi \to \pi - \xi^0$, making explicit its role as the Goldstone boson of broken time diffeomorphisms. Physically, $\pi$ parametrises fluctuations in the local clock that defines the inflationary slicing, or equivalently fluctuations in the local expansion history.

Throughout this work we assume the standard single-clock symmetry pattern of the EFT of inflation: spatial diffeomorphisms are preserved, while time diffeomorphisms are spontaneously broken and non-linearly realised by the Goldstone mode $\pi$. We do not assume exact time-translation invariance or exact de Sitter symmetry. The Wilson coefficients are therefore allowed to vary with time, as in a generic FRW background, while the effective action remains local and admits a derivative expansion. In particular, derivatives of background quantities are not required to be slow-roll suppressed. As a consequence, operator relations that hold in near–de Sitter or minimal slow-roll analyses do not reduce the counter-term basis to the same minimal set in the situations considered here.

Given this symmetry structure, the dynamics of scalar fluctuations can be organised in terms of the Goldstone mode $\pi$. During inflation it is often consistent to work in the decoupling limit, in which mixing between $\pi$ and metric fluctuations is suppressed. This limit is defined by
\begin{align}
M_{\mathrm{Pl}}^2\to\infty,
\qquad 
\dot H\to 0,
\qquad 
\text{with } \qquad
M_{\mathrm{Pl}}^2\dot H=\text{const.}
\end{align}
In this regime, gravitational fluctuations become non-dynamical and the leading scalar interactions are captured entirely by $\pi$, while gravity enters only through the background evolution.

In the decoupling limit, the action for the Goldstone mode takes the form
\begin{align}
\label{eq:exactdec}
S_{\pi}&=\int \dd^4 x\sqrt{-g}\Bigg[
\frac{M_{\mathrm{Pl}}^2}{2}R
- M_{\mathrm{Pl}}^2\big(3H^2(t+\pi)+\dot H(t+\pi)\big)
\\
&\qquad\qquad
+M_{\mathrm{Pl}}^2\dot H(t+\pi)\Big(-1-2\dot\pi+(\partial_\mu\pi)^2\Big)
+\sum_{n=2}^{\infty}\frac{M_n^4(t+\pi)}{n!}
\Big(-2\dot\pi+(\partial_\mu\pi)^2\Big)^n
\Bigg]\,, \nonumber
\end{align}
which is exact in $\pi$ within the derivative expansion implicit in the unitary-gauge operator basis. The appearance of $t+\pi$ reflects the non-linear realisation of broken time diffeomorphisms: all background quantities must be promoted to functions of $t+\pi$ in order to preserve the symmetry.

Equation~\eqref{eq:exactdec} contains an infinite tower of interactions controlled by the Wilson coefficients $M_n$ and by derivatives of background quantities such as $\dot H$. Importantly, this construction does not rely on slow roll. It remains valid as long as the time dependence of the background does not excite degrees of freedom above the EFT cutoff $\Lambda_{\rm EFT}$.

More precisely, consider a generic time-dependent coupling $M(t)$. Its variation defines a physical scale
\begin{align}
 E_{\rm var} \sim \left|\frac{\dd}{\dd t}\ln M\right|,
 \label{eq:evar}
\end{align}
which characterises the typical frequency associated with the background evolution. The EFT description remains valid provided that
\begin{align}
E_{\rm var} \ll \Lambda_{\rm EFT},
\end{align}
so that the time dependence of the couplings does not probe UV modes that have been integrated out. When this condition is satisfied, even nontrivial time dependence can be treated consistently within the low-energy effective theory.

In this paper, we will focus on the unavoidable gravitational interactions that must be present in \textit{any} inflationary theory, sometimes referred to as the {\em ``gravitational floor''}. To remain more generic, though, we will be retaining the dependence of the perturbations on the speed of sound $c_s$. This amounts to fine-tuning our theory to $M_{n\geq3}=0$, and assuming that the speed of sound defined through $2 M_2^4 \equiv - \dot{H} \mpl^2(1/c_s^2-1)$ is very slowly varying and satisfies $c_s\sim1$.

Although the action \eqref{eq:exactdec} is exact in $\pi$ within the derivative expansion of the EFT, practical computations require expanding in powers of $\pi$, as
\begin{align}
\dot H (t+\pi)
=
-\epsilon H^2
\Big(
1+ \eta H \pi
+\frac{\eta_2+\eta}{2}H^2\pi^2
+\mathcal{O}(\pi^3)
\Big),
\end{align}
where
\(
\epsilon = -\dot H/H^2,
\;
\eta = \dot\epsilon/(H\epsilon),
\;
\eta_2 = \dot\eta/(H\eta)
\)
encode the time variation of the background.

This expansion organises interactions in powers of $\pi$ and in derivatives of the Hubble scale.
To leading nontrivial order, cubic and quartic operators arise. Retaining the dominant contributions in the decoupling limit, the action up to quartic order becomes
\begin{equation}
S_{\mathrm{dec}}^{\leqslant(4)}=
\int \dd^4 x\,
a^3
\frac{\epsilon H^2M_{\mathrm{Pl}}^2}{c_s^{2}}
\Big(
1+\eta H\pi+\frac{\eta(\eta+\eta_2)}{2}H^2\pi^2
\Big)
\left[
\dot\pi^2-c_s^2\frac{(\partial_i\pi)^2}{a^2}
\right].
\label{eq:actionnon-lineardecoupling}
\end{equation}
From this expression one derives the interaction Hamiltonians. Using the operator formulation of the in-in formalism and accounting for the non-linear relation between $\dot\pi$ and its conjugate momentum, one obtains
\begin{align}
a\mathcal{H}^{(3)}_\mathrm{int} &=
     - a^4\epsilon \eta H^3 \mpl^2 
    \pi \left[
    \frac{\dot{\pi}^{ 2}}{c_s^2}
    -  \frac{(\partial_i \pi)^2}{a^2}\right]
\label{eq:H3_pi}
\equiv
\vcenter{\hbox{\begin{tikzpicture}[line width=1.pt, scale=2]
    \draw[black] (-0.2, 0) -- (0.0, 0);
    \draw[black] (0.0, 0.) -- (0.1, 0.173);
    \draw[black] (0.0, 0.) -- (0.1, -0.173);
    \node[draw, circle, fill=black, inner sep=1.5pt] at (0.0,0.0) {};
\end{tikzpicture}}}
\,,
\\
    a \mathcal{H}^{(4)}_\mathrm{int} &=
    \frac{a^4}{2}\epsilon \eta  H^4 \mpl^2  
   \pi^2 \left[
    \frac{\eta-\eta_2}{c_s^2} 
    \dot{\pi}^2
    + (\eta+\eta_2)\frac{(\partial_i \pi)^2}{a^2}
    \right]
\equiv
\vcenter{\hbox{\begin{tikzpicture}[line width=1.pt, scale=2]
\draw[black] (-0.2, 0.21) -- (0.0, 0);
\draw[black] (-0.2, -0.21) -- (0.0, 0);
\draw[black] (0.0, 0.) -- (0.2, 0.21);
\draw[black] (0.0, 0.) -- (0.2, -0.21);
\node[draw, circle, fill=black, inner sep=1.5pt] at (0.0,0.0) {};
\end{tikzpicture}}}
\,.
\end{align}
Cubic and quartic interactions are thus represented diagrammatically by vertices with three and four external legs, respectively.

In standard slow roll, one-loop corrections to the power spectrum receive contributions from cubic and quartic interactions at the same parametric order. This counting changes if the time dependence of the background is hierarchical, in the sense that higher derivatives dominate over lower ones. In the regime of interest we assume
\begin{align}
\eta_2 \gg \eta,
\end{align}
without specifying the detailed functional form of $\epsilon(t)$.

Under this hierarchy, quartic vertices scale as $\epsilon\,\eta\,\eta_2$, while cubic vertices scale as $\epsilon\,\eta$. Because a quartic interaction contributes to the one-loop two-point function with a single insertion, it generates corrections of order $\mathcal{O}(\eta\,\eta_2)$, whereas loops built solely from cubic interactions scale as $\mathcal{O}(\eta^2)$~\cite{Braglia:2025cee}. 
The quartic contribution therefore dominates parametrically.

The relevant interaction Hamiltonian reduces to
\begin{align}
    a \mathcal{H}^{(4)}_\mathrm{int} &=
    -\frac{a^4}{2}\,\epsilon \eta \eta_2  H^4 \mpl^2  
    \pi^2 \left[ \frac{\dot{\pi}^2}{c_s^2} 
    -\frac{(\partial_i \pi)^2}{a^2}
    \right]\,
\label{eq:H4_pi}
\equiv
\vcenter{\hbox{\begin{tikzpicture}[line width=1.pt, scale=2]
\draw[black] (-0.2, 0.21) -- (0.0, 0);
\draw[black] (-0.2, -0.21) -- (0.0, 0);
\draw[black] (0.0, 0.) -- (0.2, 0.21);
\draw[black] (0.0, 0.) -- (0.2, -0.21);
\node[draw, circle, fill=black, inner sep=1.5pt] at (0.0,0.0) {};
\end{tikzpicture}}}
\,.
\end{align}
To determine the scale suppressing this operator, we canonically normalisse $\pi$ using
\begin{align}
\mathcal{H}^{(2)} \supset
\, a^3
\frac{2\epsilon M_{\rm Pl}^2 H^2}{c_s^2}\dot\pi^2,
\qquad
\pi_c = \frac{2\sqrt{\epsilon} M_{\rm Pl} H}{c_s}\,\pi.
\end{align}
In terms of $\pi_c$ the quartic interaction takes the schematic form 
\begin{align}
\mathcal{H}^{(4)}_{\rm int}
\sim
\frac{1}{\Lambda_{\rm sc}^2}
\pi_c^2(\partial\pi_c)^2,
\end{align}
with
\begin{align}
\Lambda_{\rm sc}^2
\sim
\frac{16\,\epsilon\,M_{\rm Pl}^2}{\eta\,\eta_2\,c_s^2}.
\label{eq:SCScale}
\end{align}
This scale characterises when the enhanced quartic operator becomes non-perturbative. It need not coincide with the Wilsonian cutoff $\Lambda_{\rm EFT}$ of the Goldstone theory, but instead quantifies the energy at which this specific interaction loses perturbative control (see~\cite{Ballesteros:2021fsp,CarrilloGonzalez:2025fqq} for analyses of the validity of the EFT when background quantities vary in time and approach the cutoff scale $\Lambda_{\rm EFT}$.).

\subsection{Counter-terms and renormalisation}
\label{sec:counter-terms}

In order to compute loop corrections consistently, one must include all counter-terms allowed by the effective field theory in the decoupling limit and fix them through appropriate renormalisation conditions. In the EFT of inflation this procedure differs from that of a generic Poincaré-invariant quantum field theory because the background solution itself carries physical information. In particular, renormalisation must be performed so that the chosen background remains a solution of the quantum-corrected equations of motion, which amounts to cancelling tadpoles along the full in-in contour.

In unitary gauge, the set of counter-terms relevant for our purposes can be written as

\begin{align}
    \mathcal{L}_{\mathrm{c.t.}}
=&
- M_{\rm Pl}^2\,\delta\Lambda(\tau)
- \delta c(\tau)\,\delta g^{00}- \frac{\delta M_2^4 (\tau) }{2}\left( \delta g^{00}\right)^2-
     \frac{m_1^2(\tau) }{2} \left( \nabla^0 \delta g^{00} \right)^2\notag\\
      & - \frac{m_3^2(\tau) }{2} h^{ij} \left(\nabla_i \delta g^{00}\right)\left(\nabla_j \delta g^{00}\right) - \frac{\bar{M}_3^2(\tau) }{2}  \delta K^2 
     -\tilde{M}_1^3(\tau) \delta K 
      -\bar{M}_1^3(\tau) \delta g^{00} \delta K 
     \,,\label{eq:count_unitary}
\end{align}
The first two terms, proportional to $\delta\Lambda$ and $\delta c$, are tadpole counter-terms that renormalise the background equations of motion by cancelling the one-point function of the Goldstone mode. The remaining terms correspond to higher-dimensional operators that are unavoidably generated by loops in this non-renormalisable theory and must therefore be included on EFT grounds.

Restoring time diffeomorphisms and expanding in the Goldstone mode $\pi$, these counter-terms generate linear and quadratic interactions. Because the term $\delta c(t+\pi)\,\delta g^{00}$ modifies the canonical momentum of $\pi$, the interaction Hamiltonian is not simply given by minus the interaction Lagrangian. In particular, by carefully taking this into account--see e.g.~\cite{Braglia:2025cee,Braglia:2025qrb}--we can compute the interacting Hamiltonian induced by the counter-terms. The linear terms are given by:
\begin{equation}
\label{eq:Ham_count_lin}
    a \mathcal{H}_{\rm c.t.}^{(1)} =   a^4 \mpl^2\, \delta\dot{\Lambda}\, \pi-2 a^4\, \delta c\,\dot{\pi}.
\end{equation}
We note that the EFT operator $-\tilde{M}_1^3(\tau) \delta K $ also contributes to the Hamiltonian at linear order. However, in the decoupling limit, this contribution takes the form $\propto\partial^2\pi/a^2$ and reduces to a total spatial derivative. As a result, it does not contribute to correlation functions of $\pi$ and can be neglected in the computation of observables.

Next, the quadratic Hamiltonian is given by:
\begin{align}
 a \mathcal{H}_{\rm c.t.}^{(2)} 
=
a^4 M_{\rm Pl}^2
\Biggl[&
\frac{1}{2}\delta\ddot\Lambda\,\pi^2-\frac{2}{\mpl^2}  \left(\delta \dot{c}\,-\,\eta H\,\delta c \right)\,\pi\dot{\pi}
+\epsilon H^2\delta_{c_s^2}\dot\pi^2 + \delta_3(\ddot\pi)^2
+ \delta_2\frac{(\partial_i\dot\pi)^2}{a^2}\notag\\
&+ \delta_1\frac{(\partial^2\pi)^2}{a^4} + 2\delta_K\dot{\pi}\frac{\partial^2\pi}{a^2}+ \tilde{\delta}_K\left(-3H\frac{(\partial\pi)^2}{a^2}+4\frac{\partial\dot{\pi}\partial_i \pi}{a^2}+2\frac{\dot{\pi}\partial^2 \pi}{a^2}
\right)\notag\\&- \dot{\tilde{\delta}}_K\frac{\pi\,\partial^2\pi}{a^2}\Biggr].
\end{align}
Here we have retained only the leading contributions in the slow-roll expansion, keeping terms at lowest order in $\epsilon$. For convenience, we have introduced the dimensionless coefficients
\begin{equation}
    \delta_{c_s^2}\equiv\frac{2\,\delta M_2^4}{\epsilon H^2\mpl^2},\,\,\,\,\delta_3\equiv 2\frac{m_1^2}{\mpl^2},\,\,\,\, \delta_2\equiv 2\frac{m_3^2}{\mpl^2},\,\,\,\,  \delta_1\equiv 2\frac{\bar{M}_3^2}{\mpl^2},\,\,\,\,  \delta_K\equiv \frac{\bar{M}_1^3}{\mpl^2},\,\,\,\,   \tilde{\delta}_K\equiv \frac{\tilde{M}_1^3}{\mpl^2}. 
\end{equation}

We would like to rewrite this expression in a more convenient form. Let us begin with the contribution proportional to $\delta_K$. By integrating by parts, this term can be recast as~\footnote{We emphasise that the integration by parts is performed on the Hamiltonian density $\mathcal{H}$ itself, not on the combination $a\,\mathcal{H}$.}
 \begin{equation}
     a\mpl^2\, 2\delta_K\dot{\pi}\partial^2\pi=\, a \mpl^2   \left(\dot{\delta}_K+ H  \delta_K \right)(\partial_i\pi)^2+ \frac{\dd}{\dd t}\left[a \mpl^2\, \delta_K (\partial\pi)^2\right]\, ,
 \end{equation}
where we have neglected total spatial derivatives. The last term can also be neglected, as it is a total time derivative of the fields $\pi$--not its momenta--and as such it does not contribute to correlation functions of $\pi$~\cite{Braglia:2024zsl}.

A similar integration by parts can be performed to rewrite the $\tilde{\delta}_K$ terms as
\begin{align}
a\mpl^2\tilde{\delta}_K\left(-3H(\partial\pi)^2+4\partial_i\dot{\pi}\partial_i \pi+2\dot{\pi}\partial^2 \pi
\right)- a\mpl^2\dot{\tilde{\delta}}_K \pi\,\partial^2\pi=&-a\mpl^2\tilde{\delta}_K\frac{5H}{2}(\partial\pi)^2\notag\\&- \frac{\dd}{\dd t}\left[a \mpl^2\, \tilde{\delta}_K (\partial\pi)^2\right],
\end{align}
where again, we can neglect the last term which is a total time derivative of fields.

 Finally, we can simplify the $\delta_3$ term. To do so, we can use the equation of motion of the free fields, which, at lowest order in $\epsilon$ is: 
\begin{equation}
    {\rm EOM}=\ddot{\pi}+3 H \dot{\pi}+c_s^2\frac{(\partial \pi)^2}{a^2}+\mathcal{O}(\epsilon),
\end{equation}
which allows us to write,
\begin{equation}
\label{eq:ddotpisquared}
    a^3\delta_3\ddot{\pi}^2=a^3\delta_3 \left[9 H^2\dot{\pi}^2+c_s^4\frac{\left(\partial^2\pi\right)^2}{a^4}-6 H \frac{c_s^2}{a^2}\dot{\pi}\partial^2\pi\right]+\mathcal{O}({\rm EOM})\,. 
\end{equation}
In the operator formulation of the in–in formalism used here~\cite{Braglia:2024zsl}, interactions proportional to the free equations of motion do not contribute to correlation functions and can be consistently set to zero.

The last term in the square brackets of Eq.~\eqref{eq:ddotpisquared} can be further manipulated by first performing a spatial integration by parts, discarding total spatial derivatives that do not contribute to correlation functions, and then integrating by parts in time, obtaining
\begin{equation}
    -6 a^3 H \delta_3  \frac{c_s^2}{a^2}\dot{\pi}\partial^2\pi=-3Ha\left(\dot{\delta}_3+H\delta_3\right)c_s^2(\partial\pi)^2+\frac{\dd}{\dd t}\left[3 a H \delta_3 c_s^2 (\partial \pi)^2\right] 
    +\mathcal{O}(\epsilon)\,,
\end{equation}
As in the previous manipulations, total time-derivative terms can also be discarded, since they do not affect correlation functions.

Collecting all the above results, we can  write the quadratic counter-term Hamiltonian in the simplified form
\begin{equation}
    a \mathcal{H}_{\rm c.t.}^{(2)} \equiv 
    a \left(\mathcal{H}_{\rm tad}^{(2)}+
     \mathcal{H}_{\rm UV}^{(2)}\right),
 \end{equation} 
where we have defined
\begin{equation}
\label{eq:quad_tad}
    a \mathcal{H}_{\rm tad}^{(2)}\equiv  \, a^4 \mpl^2 \left[
\frac{1}{2}\delta\ddot\Lambda\,\pi^2-\frac{2}{\mpl^2}  \left(\delta \dot{c}\,-\,\eta H\,\delta c \right)\,\pi\dot{\pi}\right],
 \end{equation} 
and
\begin{align}
    a \mathcal{H}_{\rm UV}^{(2)}=  \, a^4 \mpl^2 \Biggl[& H^2\left(\epsilon\delta_{c_s^2}+9\delta_3\right) \dot{\pi}^2+ \left(\delta_1+c_s^4\delta_3\right) \frac{ \left(\partial^2\pi\right)^2}{a^4}+\delta_2  \frac{ \left(\partial_i\dot{\pi}\right)^2}{a^2}
    \notag\\& + \left(\dot{\delta}_K-3 H \dot{\delta}_3+ H \left(\delta_K-\frac{5}{2}  \tilde{\delta}_K-3 H \delta_3\right)\right)\frac{(\partial_i\pi)^2}{a^2}\Biggr].
    \label{eq:UV_quadratic}
 \end{align} 
 
 The separation between $\mathcal{H}_{\rm tad}^{(2)}$ and $\mathcal{H}_{\rm UV}^{(2)}$ reflects two  distinct roles played by quadratic counter-terms in the EFT of inflation. The terms proportional to $\delta\ddot{\Lambda}$ and $\delta\dot c$ are fixed by the requirement that tadpoles vanish at all times, ensuring that the chosen background solution remains stable under quantum corrections. By contrast, the time-dependent Wilson coefficients appearing in $\mathcal{H}_{\rm UV}^{(2)}$ are determined by the cancellation of UV divergences in correlation functions.

An important consequence of this structure is the presence of mass-like counter-terms proportional to $\pi^2$ or $\dot{\pi} \pi$, which are allowed by the symmetries of the EFT and required by background stability once loop corrections are included.
In slow-roll backgrounds, these terms play a crucial role in cancelling the secular divergences that arise in the bare one-loop correction to the two-point function. Indeed, when tadpoles and the associated quadratic counter-terms are consistently taken into account, the renormalised power spectrum remains finite and freezes on super-horizon scales, as shown explicitly in \cite{Braglia:2025cee,Braglia:2025qrb}.

\subsection{Relation between $\pi$ and $\zeta$}
At the order relevant for the loop computation, the relation between the Goldstone
mode $\pi$ and the curvature perturbation $\zeta$ can be treated linearly.
Under the time diffeomorphism $t\to t+\pi$, one finds in comoving gauge
\begin{align}
\zeta = -H\pi + \mathcal O(\pi^2),
\end{align}
where non-linear corrections arise from the gauge transformation of the metric and are
suppressed by additional powers of $\pi$ and $\epsilon$.
In the one-loop
correction generated by the quartic operator $\mathcal H^{(4)}_{\rm int}$, the diagram
contains a single interaction insertion and two external legs. At this order it is therefore
consistent to replace each external $\zeta$ by $-H\pi$, since the non-linear terms in the
$\pi\to\zeta$ mapping would contribute only to higher-point functions or to higher orders
in the loop- or $\epsilon$-expansion. As a result,
\begin{align}
\langle \zeta_{\vec p}\,\zeta_{-\vec p} \rangle
= H^2 \langle \pi_{\vec p}\,\pi_{-\vec p} \rangle
\end{align}
to the accuracy of the quartic-loop computation.

This also implies that no additional subtlety arises in the renormalisation procedure.
counter-terms are defined in the Goldstone description and cancel UV divergences
in $\langle\pi\pi\rangle$ along the in-in contour. Since the $\pi\to\zeta$ relation is
linear at the order considered, renormalising the $\pi$ two-point function automatically
renormalises the $\zeta$ correlator. Non-linear field-redefinition terms would generate
effects beyond the perturbative order retained here and therefore do not modify the
cancellation of divergences or the determination of quadratic counter-terms. In particular,
operators proportional to the free equations of motion and total time derivatives,
which are consistently discarded in the Hamiltonian, do not reappear through the
$\pi\to\zeta$ conversion. The renormalisation is thus entirely controlled at the
level of the $\pi$ action.

\section{Cancellation of UV divergences and tadpole diagrams in scale-dependent scenarios.}
\label{sec3}

We now present the contributions to the one-loop power spectrum for a general time dependence of $\epsilon$, assuming only the hierarchy $\eta_2 \gg \eta$. The corresponding in–in integrals are written in a background-independent form and regulated in the UV using dimensional regularisation, with their explicit evaluation deferred to the next section.

\subsection{Loop diagrams}

Under the hierarchy assumed above, the one-loop diagram relevant for the present analysis is generated by the quartic interaction defined in Eq.~\eqref{eq:H4_pi}. Its contribution to the power spectrum is computed using the in–in formalism and arises from closed contractions of a single quartic vertex. The corresponding diagram is shown below, and its contribution to the power spectrum is given by
\begin{align}
\label{eq:bare}
\vcenter{\hbox{\begin{tikzpicture}[line width=1. pt, scale=1.5]
    \node[draw, circle, fill, inner sep=1.2pt] (v) at (0,0) {};
    \draw[black] (-0.35,0) -- (v);
    \draw[black] (0.35,0) -- (v);
    \draw[thick, black] (0.0,0.0) arc[start angle=270, end angle=90, radius=0.2];
    \draw[thick, black](0.0,0.0) arc[start angle=-90, end angle=90, radius=0.2];
    \node[draw, circle, fill, inner sep=1.2pt] at (0,0) {};
\end{tikzpicture}}}
=&2 \Im  \frac{p^{3+\delta}\mu^{\delta}}{2\pi^2}\int\frac{\dd^{3+\delta} \vec{k}}{(2\pi)^{3+\delta}}\,     \int_{-\infty_+}^\tau \dd \tau_1 \: a^{2+\delta}(\tau_1)  \frac{H^4\mpl^2}{c_s^2} \epsilon(\tau_1)\, \eta(\tau_1)\, \eta_2(\tau_1) \,\notag\\&\pi_p^{*2}(\tau)\Bigl[\left(\pi_p'(\tau_1)\right)^2 \pi_k(\tau_1) \pi_k^*(\tau_1)+\pi_p(\tau_1)\pi_p(\tau_1) \pi_k'(\tau_1) {\pi_k^*}'(\tau_1)\notag\\&+2 \,\pi_p'(\tau_1)\pi_p(\tau_1) \pi_k'(\tau_1) \pi_k^*(\tau_1) +2 \,\pi_p'(\tau_1)\pi_p(\tau_1) \pi_k(\tau_1) {\pi_k^*}'(\tau_1)\notag\\
&-c_s^2\left(p^2+k^2\right)\pi_p^2(\tau_1) \pi_k(\tau_1) \pi_k^*(\tau_1)\Bigr].
\end{align}
In Eq.\eqref{eq:bare}, we have 
promoted the number of spatial dimensions from $3$ to $3+\delta$ using dimensional regularisation. Consistency of this procedure requires the mode functions entering the loop integral to be evaluated in $3+\delta$ dimensions as well~\cite{Senatore:2009cf}.

The integrals in Eq.~\eqref{eq:bare} diverge both in the UV and the IR. Our focus in this paper is to present a full renormalisation of the power spectrum in the UV, presenting appropriate counter-terms to remove the UV divergences. Given the debated status of IR divergences in inflation ~(see e.g.~\cite{Urakawa:2010it,Urakawa:2010kr,Giddings:2010nc,Gerstenlauer:2011ti,Tanaka:2011aj,Senatore:2012nq,Pajer:2013ana} for related works, and~\cite{Seery:2010kh} for a review) we will be simply regulating them with a comoving cut-off $\Lambda_{\rm IR}$, allowing us to track their size. We will, however, postpone their interpretation to future works.

With this in mind, our in-in integrals take schematically the following 
form\footnote{We refer the reader to Section~3.1 of~\cite{Braglia:2025cee} for the details of the calculation of the integral over the momentum in dim-reg.}
\begin{equation}
    \int\frac{\dd^{3+\delta} \vec{k}}{(2\pi)^{3+\delta}} f= \left[\int\frac{\dd^{3+\delta} \vec{k}}{(2\pi)^{3+\delta}}\,f\right]_{\rm IR} +\left[\int\frac{\dd^{3+\delta} \vec{k}}{(2\pi)^{3+\delta}}\,f\right]_{\rm UV}+\left[\int\frac{\dd^{3+\delta} \vec{k}}{(2\pi)^{3+\delta}}\,f\right]_{\rm finite}\,,
\end{equation}
where $f=f\left(\vec{k},\,\tau_1\,\right)$ denotes the loop-momentum integrand at fixed internal time $\tau_1$, and the first and second terms 
schematically denote this parts of the integrand that, after integration over the loop momentum, give ries to IR divergences and $\delta^{ -1}$ UV poles, respectively, while the last one is always finite.

The existence of an adiabatic regime continuously related to the Bunch-Davies vacuum imposes that 
the divergent parts of the integral arise only from two of the contractions of the mode functions in Eq.~\eqref{eq:bare}.
We prove this statement, independent of the precise inflationary expansion history, in Appendix~\ref{appC}.
Focusing on the UV part of the integral, which is the one that we will eventually renormalise, the divergent part of the integral is given by:
\begin{align}
\left[\vcenter{\hbox{\begin{tikzpicture}[line width=1. pt, scale=1.5]
    \node[draw, circle, fill, inner sep=1.2pt] (v) at (0,0) {};
    \draw[black] (-0.35,0) -- (v);
    \draw[black] (0.35,0) -- (v);
    \draw[thick, black] (0.0,0.0) arc[start angle=270, end angle=90, radius=0.2];
    \draw[thick, black](0.0,0.0) arc[start angle=-90, end angle=90, radius=0.2];
    \node[draw, circle, fill, inner sep=1.2pt] at (0,0) {};
\end{tikzpicture}}}
\right]_{\rm UV}
=&2 \Im  \frac{p^{3+\delta}\mu^{\delta}}{2\pi^2}     \int_{-\infty_+}^\tau \dd \tau_1 \: a^{2+\delta}(\tau_1)  \frac{H^4\mpl^2}{c_s^2} \epsilon(\tau_1)\, \eta(\tau_1)\, \eta_2(\tau_1) \,\notag\\&\times\pi_p^{*2}(\tau)\Bigl[{\pi_p'}^2(\tau_1) -c_s^2 p^2\pi_p^2(\tau_1)\Bigr]\left[\int\frac{\dd^{3+\delta} \vec{k}}{(2\pi)^{3+\delta}}\,\lvert\pi_k(\tau_1)\rvert^2\right]_{\rm UV}\,,\label{eq:UV}
\end{align}
where, for later convenience, we have switched the order of the integration over time and momentum compared to Eq.~\eqref{eq:bare}.

\subsection{Cancellation of UV divergences}
We now introduce a suitable combination of counter-terms to remove the UV divergences and render the resulting power spectrum finite, up to infrared divergences that we will only comment on later. More precisely, we impose that the cancellation of UV divergences holds at all times during inflation. 

This requirement is non-trivial due to the time dependence of the interaction couplings. After performing the in-in integral, this leads to a UV pole in $\delta$ whose coefficient depends non-trivially on both the finite time $\tau$ and the power-spectrum wave-number $p$. In contrast to scale-invariant scenarios, this dependence is more complicated than a simple power of $x=-p\tau$.
As a consequence, a suitable choice for the functional form of the Wilson coefficients of the counter-terms is required.

Using the interactions in Eq.~\eqref{eq:UV_quadratic}, we can write the contribution of the counter-terms to the power spectrum as:
\begin{align}
\vcenter{\hbox{\begin{tikzpicture}[line width=1. pt, scale=2]
    \draw[black] (-0.35,0) -- (-0.06,0);
    \draw[black] (0.06,0) -- (0.35,0) ;
\node[draw, circle, minimum size=7pt, inner sep=0pt,
          path picture={\draw[line width=1pt] 
            (path picture bounding box.south west) -- (path picture bounding box.north east)
            (path picture bounding box.north west) -- (path picture bounding box.south east);}
         ] (X) at (0,0) {};       
\end{tikzpicture}}}
=&-4 \mpl^2 \Im  \frac{p^{3+\delta}\mu^{\delta}}{2\pi^2}\,     \int_{-\infty_+}^\tau \dd \tau_1 \: a^{2+{\delta}}(\tau_1) \pi_p^{*2}(\tau)\,\Biggl[ \left(\epsilon  \delta_{c_s^2}(\tau_1)+9 \delta_3(\tau_1)\right)H^2 {\pi_p'}^{2}(\tau_1)\notag\\
&+\left(-\frac{5}{2}H\tilde{\delta}_K(\tau_1)+\dot{\delta}_K(\tau_1)+H\delta_K(\tau_1)-3H\left(\dot{\delta}_3(\tau_1) + H \delta_3(\tau_1)\right)c_s^2 \right)  p^2 \pi^2_p(\tau_1)\notag\\
&+\left(\delta_1(\tau_1)+c_s^4 \delta_3(\tau_1)\right) \frac{p^4}{a^2}{\pi_p}^{2}(\tau_1)+ \delta_2(\tau_1) \frac{p^2}{a^2}{\pi_p'}^{2}(\tau_1)\Biggr].
\label{eq:ct}
\end{align}
The structure of this integral is very similar to that of Eq.~\eqref{eq:UV}. We can therefore choose the Wilson functions such that the sum of Eqs.~\eqref{eq:UV} and~\eqref{eq:ct} is UV finite. To this end, it is convenient to parametrise them as $\delta_i(\tau)\equiv C_i\,\epsilon(\tau)\,\eta(\tau)\,\eta_2(\tau)$, where the $C_i$ are constants. This leads to the following solution:
\begin{align}
C_1&=-C_3 c_s^4\,,\quad\quad
C_2=0\,,\quad\quad
C_K=3 H c_s^2 C_3\,, \\
-\frac{5}{2}H \tilde{C}_K&=-H^4\mpl^2\left[\int\frac{\dd^{3+\delta}k}{(2\pi)^{3+\delta}}\lvert\pi_k(\tau_1)\rvert^2\right]_{\rm UV}\,,\nonumber\\
\epsilon H^2\, C_{c_s^2}+9H^2 C_3&=\frac{H^4\mpl^2}{c_s^2}\left[\int\frac{\dd^{3+\delta}k}{(2\pi)^{3+\delta}}\lvert\pi_k(\tau_1)\rvert^2\right]_{\rm UV}\,.\nonumber
\end{align}
At this point, we see that some of the counter-terms are redundant. In particular, there are two possible choices. In the first one, $\tilde{C}_K$, $C_1$, $C_3$, and $C_K$ are non-vanishing, while $C_{c_s^2}$ and $C_2$ are set to zero. In the second one, $C_{c_s^2}$ and $\tilde{C}_K$ are non-vanishing, while $C_1$, $C_2$, $C_3$, and $C_K$ are set to zero. We choose the second option, which is more minimal, and reads
\begin{align}
\label{eq:solved_counter-terms}
C_1\,&=C_2\,=C_3\,=C_K\,=0\,,\\
 \tilde{C}_K&=\frac{2}{5} H^3\mpl^2\left[\int\frac{\dd^{3+\delta}k}{(2\pi)^{3+\delta}}\lvert\pi_k(\tau_1)\rvert^2\right]_{\rm UV}\,,\nonumber\\
\epsilon H^2\,C_{c_s^2}&=\frac{H^4\mpl^2}{c_s^2}\left[\int\frac{\dd^{3+\delta}k}{(2\pi)^{3+\delta}}\lvert\pi_k(\tau_1)\rvert^2\right]_{\rm UV}\,.\nonumber
\end{align}
In particular, we see that $\tilde{C}_K\neq0$ for both choices. Notably, this operator is often assumed to be degenerate with other operators in seminal works on the EFT of inflation~\cite{Cheung:2007st,Noumi:2012vr}.
At this stage, no particular time dependence of the slow-roll parameters is assumed. The result is therefore valid independently of the specific form of the mode functions for the field $\pi$. Moreover, since the non-zero counter-terms are of order $\delta^{-1}$, the $\mathcal{O}(\delta)$ contributions to Eq.~\eqref{eq:ct} generate finite terms that must be taken into account when quoting the renormalised power spectrum. In this sense, our approach corresponds to a minimal-subtraction prescription. We fix the constants $C_i$ so that the $C_i,\mathcal{O}(\delta^0)$ contributions cancel the UV poles proportional to $1/\delta$. Within this prescription, the finite contributions induced by the counterterms are then obtained from the $C_i\,\mathcal{O}(\delta^1)$ terms in Eq.~\eqref{eq:ct}, multiplied by the corresponding $1/\delta$ coefficients. As usual, however, additional finite local counterterms could be added and would correspond to a different renormalisation prescription. These finite terms should ultimately be fixed by renormalisation conditions or by matching to observables, i.e. by specifying the renormalised EFT coefficients at a reference scale.

\subsection{Tadpole diagrams and their contribution to the power spectrum}
Although cubic interactions lead to sub-leading bare loop contributions to the power spectrum in the regime of interest, they do contribute to the leading one-point function of the Goldstone boson $\pi$, and hence to that of the curvature perturbation $\zeta$, through tadpole diagrams. A non-vanishing expectation value of $\zeta$ would backreact on the background dynamics, effectively shifting the scale factor as $a(t)\mapsto a(t)\,e^{\langle\zeta(t)\rangle}$. As a result, the in–in formalism would need to be reformulated around this shifted background, and additional non–one-particle-irreducible contributions to the power spectrum would have to be taken into account.

A simpler and more convenient way to account for such backreaction effects is to absorb them into linear counter-terms~\cite{Pimentel:2012tw}. For the cubic interactions in Eqs.~\eqref{eq:H3_pi}, this procedure was carried out in~\cite{Braglia:2025cee,Braglia:2025qrb}, and we briefly summarise the relevant results here.

The contributions from the tadpole diagrams read:
\begin{align}
  \vcenter{\hbox{\begin{tikzpicture}[line width=1. pt, scale=2]
    \draw[black] (0,-0.2) -- (0,0);
    \draw[thick, black] (0.0,0.0) arc[start angle=270, end angle=90, radius=0.15];
    \draw[thick, black](0.0,0.0) arc[start angle=-90, end angle=90, radius=0.15];
    \node[draw, circle, fill, inner sep=1.5pt]  at (0.0,0.0) {};    
\end{tikzpicture}}} =&\,
2 {\rm Im}\, \mu^\delta \pi^*_p(\tau)\int\dd\tau_1 a^4(\tau_1)\pi_p(\tau_1) \epsilon(\tau_1)\eta(\tau_1) H^3\mpl^2\int \frac{\dd^{3+\delta} \vec{k}}{(2\pi)^{3+\delta}}\frac{\lvert\dot{\pi}_k(\tau_1)\rvert^2}{c_s^2}\label{eq:tad_1}\\
&+2 {\rm Im}\, \mu^\delta \pi^*_p(\tau)\int\dd\tau_1 a^4(\tau_1)\dot{\pi}_p(\tau_1) \epsilon(\tau_1)\eta(\tau_1) H^3\mpl^2\int \frac{\dd^{3+\delta} \vec{k}}{(2\pi)^{3+\delta}}\left[\frac{\dot{\pi}_k(\tau_1)\pi^*_k(\tau_1)}{c_s^2}+{\rm c.c.}\right]\notag
,\\
& -2 {\rm Im}\, \mu^\delta \pi^*_p(\tau)\int\dd\tau_1 a^4(\tau_1)\pi_p(\tau_1) \epsilon(\tau_1)\eta (\tau_1)H^3\mpl^2\int \frac{\dd^{3+\delta} \vec{k}}{(2\pi)^{3+\delta}}\left(\frac{k}{a}\right)^2\vert\pi_k(\tau_1)\rvert^2.
\label{eq:tad_2}
\end{align}
Note that these momentum integrals are convergent both in the IR and in the UV and therefore no $\delta^{-1}$ pole arises in dimensional regularisation, see App.~\ref{appC} for more details.

The linear counter-terms, on the other hand,
contribute to the 1-point function through Eq.~\eqref{eq:Ham_count_lin} as
 \begin{equation}
\langle\pi_{\vec{p}}(\tau)\rangle'_{\rm c.t.}\,=\,
\vcenter{\hbox{\begin{tikzpicture}[line width=1. pt, scale=2]
    \draw[black] (0,-0.3) -- (0,-0.06);
\node at (0, 0.2) {$\delta\dot{\Lambda}$};
\node[draw,  minimum size=7pt, inner sep=0pt,
          path picture={\draw[line width=1pt] 
            (path picture bounding box.south west) -- (path picture bounding box.north east)
            (path picture bounding box.north west) -- (path picture bounding box.south east);}
         ] (X) at (0,0) {};   
\end{tikzpicture}}}+
\vcenter{\hbox{\begin{tikzpicture}[line width=1. pt, scale=2]
    \draw[black] (0,-0.3) -- (0,-0.06);
\node at (0, 0.2) {$\delta c$};
\node[draw,  minimum size=7pt, inner sep=0pt,
          path picture={\draw[line width=1pt] 
            (path picture bounding box.south west) -- (path picture bounding box.north east)
            (path picture bounding box.north west) -- (path picture bounding box.south east);}
         ] (X) at (0,0) {};   
\end{tikzpicture}}}
= -2 {\rm Im}\, \mu^\delta \pi^*_p(\tau)\int\dd\tau_1 a^4(\tau_1)\left[\mpl^2 \delta\dot{\Lambda}(\tau_1) \pi_p(\tau_1) -2 \frac{\delta c(\tau_1)}{c_s^2}\dot{\pi}_p(\tau_1) \right].
\end{equation}
We now require that the linear counter-terms cancel the tadpoles as follows:
\begin{equation}
\label{eq:tadpole_cancellation}
\langle\pi_{\vec{p}}(\tau)\rangle'_{\rm ren}=\langle\pi_{\vec{p}}(\tau)\rangle'_{\rm bare}+\langle\pi_{\vec{p}}(\tau)\rangle'_{\rm c.t.}
=
\vcenter{\hbox{\begin{tikzpicture}[line width=1. pt, scale=2]
    \draw[black] (0,-0.2) -- (0,0);

    \draw[thick, black] (0.0,0.0) arc[start angle=270, end angle=90, radius=0.15];
    \draw[thick, black](0.0,0.0) arc[start angle=-90, end angle=90, radius=0.15];

    \node[draw, circle, fill, inner sep=1.5pt]  at (0.0,0.0) {};    
\end{tikzpicture}}}
+
\vcenter{\hbox{\begin{tikzpicture}[line width=1. pt, scale=2]
    \draw[black] (0,-0.3) -- (0,-0.06);

\node[draw,  minimum size=7pt, inner sep=0pt,
          path picture={\draw[line width=1pt] 
            (path picture bounding box.south west) -- (path picture bounding box.north east)
            (path picture bounding box.north west) -- (path picture bounding box.south east);}
         ] (X) at (0,0) {};  
         
\node at (0, 0.2) {$\delta\dot{\Lambda}$};
\end{tikzpicture}}}+
\vcenter{\hbox{\begin{tikzpicture}[line width=1. pt, scale=2]
    \draw[black] (0,-0.3) -- (0,-0.06);

\node[draw,  minimum size=7pt, inner sep=0pt,
          path picture={\draw[line width=1pt] 
            (path picture bounding box.south west) -- (path picture bounding box.north east)
            (path picture bounding box.north west) -- (path picture bounding box.south east);}
         ] (X) at (0,0) {};  
         
\node at (0, 0.2) {$\delta c$};
\end{tikzpicture}}}=0\,,
\end{equation}
which uniquely fixes
\begin{align}
\label{eq:delta_Lambda}
   \delta{\dot{\Lambda}}(\tau_1)=&\epsilon(\tau_1)\eta(\tau_1) H^3 \int\frac{\dd^{3+\delta} \vec{k}}{(2\pi)^{3+\delta}}\left[\frac{\lvert\dot{\pi}_k(\tau_1)\rvert^2}{c_s^2}-\left(\frac{k}{a}\right)^2\vert\pi_k(\tau_1)\rvert^2\right],\\
\label{eq:delta_c}
   \delta{c}(\tau_1)=&-\frac{\epsilon(\tau_1)\eta(\tau_1)}{2 } H^3 \mpl^2 \int\frac{\dd^{3+\delta} \vec{k}}{(2\pi)^{3+\delta}}\left[\dot{\pi}_k(\tau_1)\pi^*_k(\tau_1) + {\rm c.c.} \right].
\end{align}

Thanks to the non-linearly realised symmetries of the EFT, however, the same operators producing the linear counter-terms, also generate quadratic counter-terms, which are in turn uniquely fixed by the condition for the tadpole cancellation. We derived the corresponding quadratic Hamiltonian in~\eqref{eq:quad_tad}. However, from the explicit form of $\delta\dot{\Lambda}$ and $\delta c$, we see that in our regime of validity where $\eta_2\gg\eta$, we have\footnote{We note that to compute $\delta\ddot{\Lambda}$ and $\delta\dot{c}$, one should also derive the mode functions inside the momentum integrals in Eqs.~\eqref{eq:delta_Lambda} and \eqref{eq:delta_c} with respect to time. For simplicity, we neglect them in the following. This is completely fine for de Sitter mode functions--which we will be using in the concrete examples in the next Section--for which the momentum integrals are time-independent, so their time-derivative is zero. However, these terms may play a role in more general scenarios.}  $\delta\ddot{\Lambda}\sim H \eta_2\,\delta\dot{\Lambda}$ and $\delta\dot{c}\sim H \eta_2\,\delta c$, so we can neglect the last term in~\eqref{eq:quad_tad},and these interactions contribute to the 1-loop power spectrum as:
\begin{align}
{\hbox{\begin{tikzpicture}[line width=1. pt, scale=2]
    \draw[black] (-0.35,0) -- (-0.06,0);
    \draw[black] (0.06,0) -- (0.35,0) ;
    \node at (0, 0.2) {$\delta\ddot{\Lambda}$};
\node[draw,  minimum size=7pt, inner sep=0pt,
          path picture={\draw[line width=1pt] 
            (path picture bounding box.south west) -- (path picture bounding box.north east)
            (path picture bounding box.north west) -- (path picture bounding box.south east);}
         ] (X) at (0,0) {};  
\end{tikzpicture}}}+{\hbox{\begin{tikzpicture}[line width=1. pt, scale=2]
    \draw[black] (-0.35,0) -- (-0.06,0);
    \draw[black] (0.06,0) -- (0.35,0) ;
    \node at (0, 0.2) {$\delta\dot{c}$};
\node[draw,  minimum size=7pt, inner sep=0pt,
          path picture={\draw[line width=1pt] 
            (path picture bounding box.south west) -- (path picture bounding box.north east)
            (path picture bounding box.north west) -- (path picture bounding box.south east);}
         ] (X) at (0,0) {};  
\end{tikzpicture}}} & =-2 \Im  \frac{p^{3+\delta}\mu^{\delta}}{2\pi^2}\,     \int_{-\infty_+}^\tau \dd \tau_1 \: a^{4+\delta}(\tau_1) \mpl^2\, \delta\ddot{\Lambda}\,(\tau_1)\,\pi_p^{*2}(\tau)\,\pi_p^2(\tau_1) \notag\\+&\,8 \Im  \frac{p^{3+\delta}\mu^{\delta}}{2\pi^2}\,     \int_{-\infty_+}^\tau \dd \tau_1 \: a^{3+\delta}(\tau_1) \,\delta\dot{c}(\tau_1) \,\pi_p^{*2}(\tau)\,\pi_p(\tau_1)\,{\pi_p'}(\tau_1) \notag.
\end{align}
We will compute these contributions for explicit models in the next Section.
\section{Small-amplitude primordial features}
\label{sec4}

After the generic conclusions of the previous section, we now turn to some concrete model realisations where all calculations can be performed.
For this matter, we consider that $\epsilon$ acquires a non-trivial time dependence through:
\begin{equation}
\label{eq:eps_time_dep}
\epsilon(\tau)
=
\epsilon_0+\Delta\epsilon(\tau)
=
\epsilon_0\left[
1+A f(\tau) \right]\,,
\end{equation}
where $f(\tau)$ is a function of time that will be made explicit later.
The overall amplitude of $\epsilon$, i.e. $\epsilon_0$, is a small parameter as required by the decoupling limit that we have assumed.
Therefore, we should still work at leading non-vanishing order in $\epsilon_0$ and we can neglect the variations of $H$ when it is already multiplied by $\epsilon$.
However, the time dependence of $\epsilon$ is transferred into its own derivatives, $\eta$ and $\eta_2$.
We will make the further assumption that the parameter controlling the size of the primordial feature, $A$, is also very small, \textit{i.e.} $A\ll 1$.
Under this regime, which is the one relevant to fit CMB residuals with primordial features~\cite{Planck:2018jri,Braglia:2022ftm}, we can simply treat $\Delta\epsilon(\tau)$ as a time-dependent interaction that perturbs the free theory introduced above--see Appendix~\ref{appA} for details--and use the de Sitter mode functions in our calculations. 
In $3+\delta$ spatial dimensions, they take the form:
\begin{align}
\label{eq:pi_d_exp}
	\pi_k (\tau)\underset{\delta\to0}{=}&
    -i \frac{1}{2\sqrt{ c_s\epsilon}} \frac{ 1    }{\mpl k^{3/2}}\left(\frac{H}{ \mu}\right)^{\delta/2} ( 1+i c_s k \tau)e^{-i 
		c_s k \tau } \Biggl[1+\frac{\delta}{2}\Biggl(  \ln (-\tau)\\
        &+ \frac{1}{1+ i c_s k \tau} - \frac{1- i c_s k \tau}{2(1+i c_s k \tau)} e^{2 i c_sk \tau} \left(- \pi i +  \mathrm{Ei}(-2 i c_s k \tau) \right)\Biggr)\Biggr]+ \mathcal{O}(\delta^2), \nonumber
\end{align}
where the expansion has been carried out to linear order in $\delta$, which is sufficient to capture the divergent and finite contributions in dimensional regularisation.

Furthermore, we have the following relations:
\begin{equation}
\eta
=
 A \frac{\dot f}{H}
+\mathcal{O}(A^2)\,, \quad
\eta_2
=
\frac{\ddot f}{\dot{f} H}+\mathcal{O}(\epsilon_0)
+\mathcal{O}(A) \quad \implies \quad \eta_2 \gg \eta \,.
\end{equation}
This hierarchy, resting on the smallness of $A$, ensures the condition $\eta_2 \gg \eta$ that we have assumed in Sec.~\ref{sec2}, and justifies why the quartic interaction dominates over the cubic ones at one-loop order.
In practice, the time-dependent background functions appearing in the in-in integrals for the renormalised primordial power spectrum are always of the form $a^n H^m \epsilon \eta \eta_2 $, which, under our regime of assumptions, reads
\begin{equation}
    a^n H^m \epsilon \eta \eta_2 = H^{m-n-2} \epsilon_0 A \times (-\tau)^{-n} \ddot{f}|_\tau \,,
\end{equation}
where we have factored out the term that can consistently be considered a constant.

In the following subsections, we will focus on two  concrete examples of the function $f(\tau)$, compute the 1-loop renormalised power spectrum at the end of inflation, and discuss its observability. Technical results, as well as plots for the time evolution of the loop corrections, are reported in Appendix~\ref{appB}.

\subsection{Resonant features}

\begin{figure}
        \centering
        \includegraphics[width=\textwidth]{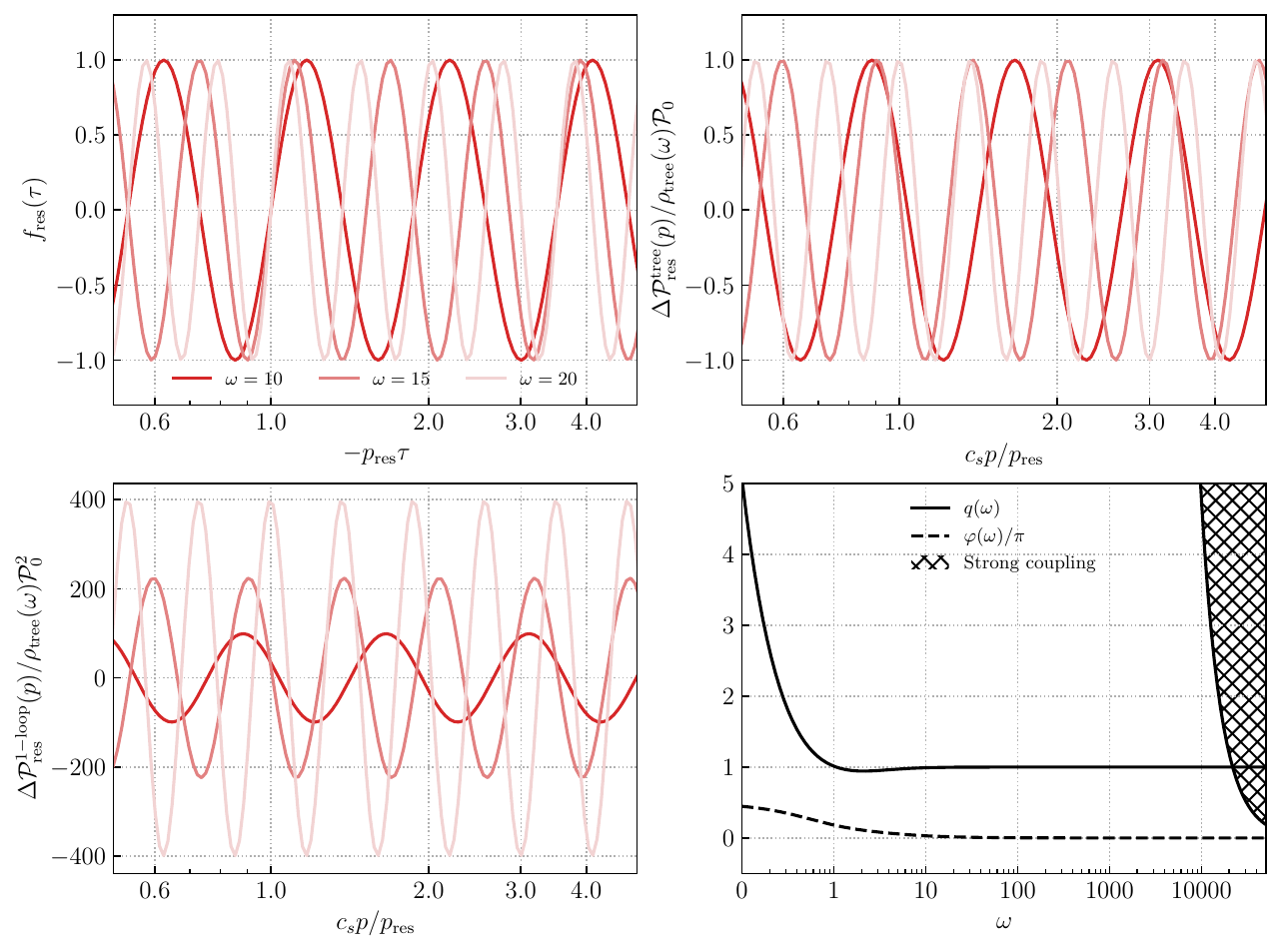}
        \caption{[Top-Left] Time-dependent correction to the first SR parameter $\epsilon$ for the resonant model. Tree-level [Top-Right] and 1-loop [Bottom-Left] corrections to the scalar power spectrum at the end of inflation. In these last 2 panels, we normalise the spectra by the maximum amplitude of the correction to the tree-level power spectrum. [Bottom-Right] Amplitude $q(\omega) = \rho_{\rm 1\text{-}loop}(\omega)/[\mathcal{P}_0 \omega^2 \rho_{\rm tree}(\omega)]$ (solid lines), and phase difference $\varphi(\omega)$ (dashed lines) as a function of $\omega$. The hatched region signals strong coupling.
        }
        \label{fig:resonant}
\end{figure}

\paragraph{Setup.}
We start by considering resonant features in the inflationary dynamics, first introduced in covariant models with an oscillating inflaton potential~\cite{Chen:2008wn,Flauger:2009ab,Flauger:2010ja,McAllister:2008hb,Chen:2010bka}, which we model
as Eq.~\eqref{eq:eps_time_dep} with 
\begin{equation}
\label{eq:eps_res}
A
=
A_{\rm res} \,, \quad f(\tau)=f_{\rm res}(\tau) \equiv 
\sin\!\left(\omega\ln\frac{\tau}{\tau_{\rm res}}\right)
\,,
\end{equation}
from which we can define the reference scale
$p_{\rm res}\equiv-\tau_{\rm res}^{-1}$.
See the left panel of Fig.~\ref{fig:resonant} for the explicit time dependence of $f_{\rm res}(\tau)$, depending on the dimensionless frequency $\omega$.
This background explicitly breaks continuous time translations and dilatations, while preserving an exact discrete remnant. Since the oscillation depends only on $\ln(-\tau)$, the background is invariant under the discrete rescaling
\begin{equation}
\tau \;\to\; e^{2\pi n/\omega}\,\tau\,, \qquad n\in\mathbb{Z},
\end{equation}
which corresponds to a discrete subgroup of time translations,
$t\to t+2\pi/(\omega H)$. Within the EFT of inflation, where time diffeomorphisms are non-linearly realised by the Goldstone mode $\pi$, this symmetry breaking pattern reduces the continuous shift symmetry of $\pi$ to a discrete one,
$\pi\to\pi-2\pi/(\omega H)$. As a result, \textit{all} EFT coefficients are constrained to be periodic functions of time, or equivalently to admit a Fourier expansion in harmonics of the frequency $\omega$, rather than arbitrary smooth functions~\cite{Behbahani:2011it}.
In particular, this applies to the coefficients of the counter-terms of the theory.
This residual discrete symmetry leads to logarithmic oscillations and discrete scale invariance in correlation functions, $k\to e^{2\pi/\omega}k$, which underlie the characteristic resonant features in the power spectrum and non-Gaussianity, while the EFT remains local, unitary, and organised by the non-linearly realised breaking of time translations.

\paragraph{Tree-level scale dependence.}
Our goal is to compute the renormalised primordial power spectrum of the curvature perturbation $\zeta$ in the presence of the resonant feature \eqref{eq:eps_res}.
At tree level, this can be done either by solving the Sasaki-Mukhanov equation with the full time dependence induced by $\Delta\epsilon(\tau)$, or perturbatively using the in-in formalism. 
The two approaches should agree for $A_{\rm res}\ll1$, as shown explicitly in~\cite{Chen:2015dga}. 
In what follows, we adopt the perturbative approach, which is naturally suited for extensions beyond tree level.
Within this framework, the leading correction to the power spectrum can be computed analytically, see App.~\ref{appA} for more details.
At late times, $\tau\to0$, the fractional correction takes the closed form
\begin{align}
\label{eq:result_integral_P_res}
\frac{\Delta\mathcal{P}_{\rm res}^{\rm tree}(p)}{\mathcal{P}_0}
&= \rho_{\rm tree}(\omega) \cos\left[\omega\ln\left(\frac{2 c_s p}{p_{\rm res}}\right) + \phi_{\rm tree}(\omega)\right],
\end{align}
where $\mathcal{P}_0 = H^2/(8 \pi^2 \epsilon_0 c_s \mpl^2)$ is the scale-invariant treel-level power spectrum amplitude and
\begin{align}
    \rho_{\rm tree}(\omega)&= A_{\rm res}\sqrt{\frac{\pi\omega/2}{\tanh{\pi\omega/2}}}\, ,
    \\
    \phi_{\rm tree}(\omega)&=\mathrm{arg}\left[(\omega+i)^2\Gamma(-i\omega)\right].
\end{align}
This expression displays the characteristic $\cos(\omega\ln p)$ modulation of resonant features.
For $A_{\rm res}\ll1$ it agrees with the numerical solution of the Sasaki-Mukhanov equation, as illustrated in Fig.~\ref{fig:Pk_tree_res} in App.~\ref{appA}.
At larger amplitudes, deviations signal the breakdown of the linear expansion in $A_{\rm res}$, while preserving the same oscillatory scale dependence, which is guaranteed by the residual discrete shift symmetry.

\paragraph{Loop-level scale dependence.}
We have already tackled in full generality the UV divergences of the bare loops.
We have shown how to cancel the $\delta^{-1}$ poles with local counter-terms respecting the EFT symmetries.
In particular, note that the $\delta_i(\tau)$ coefficients that we have solved for in Eq.~\eqref{eq:solved_counter-terms} are proportional to $\epsilon \eta \eta_2 \propto \ddot{f} \propto f$, thus preserving the discrete time shift symmetry.

What remains is the sum of all finite contributions, at order $\delta^0$.
For each contribution, we compute the integrals including UV-finite terms by following the procedure\footnote{This procedure is reminiscent of, but distinct from, the so-called method of regions, see e.g.~\cite{Beneke:1997zp,Smirnov:1999bza} and~\cite{Beneke:2023wmt,Beneke:2026rtf,Beneke:2026ksj} for applications to cosmological correlation functions. In a method-of-regions expansion of late-time cosmological correlators, the relevant integration variables are expanded according to their scaling with a late-time parameter, which can involve both vertex times and loop momenta. Here, instead, we only isolate the UV-sensitive part of the loop-momentum integrand and perform the in-in time integrals exactly. For a discussion of the differences between the two approaches, see also Sec.~2 of Ref.~\cite{Ballesteros:2024cef}.} recently developed in Ref.~\cite{Ballesteros:2024cef}, subsequently used in\footnote{See also Ref.~\cite{Ballesteros:2025nhz} for a recent extension of this methodology.}~\cite{Braglia:2025cee,Braglia:2025qrb}, i.e.:
\begin{itemize}
    \item We use the mode functions~\eqref{eq:pi_d_exp} and expand the integrand at order $\delta^1$;
    \item We perform the time integration first;
    \item We expand the integrand for large internal momenta $k$ as we are only interested in the UV divergence part of the loop, and perform the integral over it.
    This brings a pole $\delta^{-1}$ that multiplies the overall $\delta^{1}$ factor, thus leaving \textit{UV-finite contributions} at order $\delta^0$.
\end{itemize}
We simply quote the final result, at late times, $\tau \rightarrow 
0$, for the fractional correction at one-loop level:
\begin{align}
\label{eq:result_integral_P_res_loop}
\frac{\Delta\mathcal{P}_{\rm res}^{\rm 1\text{-}loop}(p)}{\mathcal{P}_0^2}
&= \rho_{\rm 1\text{-}loop}(\omega) \cos\left[\omega\ln\left(\frac{2 c_s p}{p_{\rm res}}\right) + \phi_{\rm 1\text{-}loop}(\omega)\right]+\frac{\Delta \mathcal{P}_{\rm res}^{\rm logs} (p)}{\mathcal{P}_0^2}\,,
\end{align}
where $\Delta \mathcal{P}_{\rm res}^{\rm logs} (p)$ includes various logarithmic terms such as $\ln (H/\mu)$ and $\ln(\Lambda_{\rm IR}/p)$, and
\begin{align}
    \rho_{\rm 1\text{-}loop}(\omega)&= \rho_{\rm tree}(\omega) \times \omega^2 \times q(\omega) \quad \text{with} \,\,q(\omega) \underset{\omega \rightarrow \infty}{\longrightarrow} 1\,,
    \\
    \phi_{\rm 1\text{-}loop}(\omega)&= \phi_{\rm tree}(\omega) +\varphi(\omega)  \quad \quad \quad \text{with} \,\,\varphi(\omega) \underset{\omega \rightarrow \infty}{\longrightarrow} 0\,.
\end{align}
The exact expressions for $\Delta \mathcal{P}_{\rm res}^{\rm logs} (p), q(\omega), \varphi(\omega)$ can be found in App.~\ref{appB}, see also Fig.~\ref{fig:resonant} below.
As expected, the discrete time-translation symmetry of the theory manifests itself in enforcing the scale dependence of the power spectrum, even at loop level, to be invariant under discrete $\ln p$-translations of factors of $2 \pi / \omega$.
Note also that the presence of the primordial feature automatically regulates the late-time behaviour of the one-loop power spectrum, and therefore none of the individual contributions contains secular divergences, as opposed to the scale-invariant case~\cite{Braglia:2025cee,Braglia:2025qrb}. 
We show in Fig.~\ref{fig:resonant} the time dependence of $\epsilon$ for this resonant model and a few values of $\omega$, as well as the resulting tree-level and loop-level scale-dependent corrections to the primordial power spectrum.
Clearly, the loop correction grows more quickly than the tree-level one with $\omega$, as can also be read from the expression of $\rho_{\rm 1 \text{-} loop}(\omega)$, thus pointing to a physically meaningful perturbativity criterion.
Moreover, we realise that both corrections are in phase in the large $\omega$ limit.

Finally, let us comment on the logarithmic term proportional to $\ln(\Lambda_{\rm IR}/p)$. This term seems to break the discrete time-translation symmetry respected by the resonant model. However, it originates from our choice of IR prescription, namely a comoving cutoff on the loop momentum, which itself does not preserve the discrete rescaling symmetry. For example, if the IR part of the integral were regulated dimensionally, with an IR regulator $\delta_{\rm IR}$ independent from the UV one, the cutoff logarithm would be replaced schematically by
$\delta_{\rm IR}$ as $\ln(\Lambda_{\rm IR}/p)\mapsto-1/\delta_{\rm IR}-\ln\left(H/\mu_{\rm IR}\right)$
up to scheme-dependent finite terms. The coefficient of the corresponding $1/\delta_{\rm IR}$ pole respects the discrete time-translation symmetry which, therefore, does not develop any anomaly.
A complete treatment of the finite terms associated with the IR dimensional regulator is beyond the scope of the present work and will be addressed elsewhere.

\paragraph{Perturbativity.}
The notion of perturbativity in scale-dependent scenarios is subtle.
First, we have used a perturbation theory where the whole of the scale dependence is treated perturbatively, even at tree level--see Appendix~\ref{appA}.
Therefore, what we should ask is rather that the loop-level corrections $\Delta \mathcal{P}_{\rm res}^{\rm 1\text{-}loop}(p)$ be smaller than the tree-level ones $\Delta \mathcal{P}_{\rm res}^{\rm tree}(p)$, which is more restrictive than asking them to be smaller than the total tree-level\footnote{We note that the finite parts of our results are scheme dependent. A more precise estimate would require specifying renormalisation conditions, or equivalently fixing the finite local counterterms/renormalised Wilson coefficients through a matching to observables. The results of our perturbativity analysis should therefore be understood as order-of-magnitude estimates within our minimal-subtraction prescription, and may receive $\mathcal{O}(1)$ corrections in the model parameters once such a matching prescription is specified.} $\mathcal{P}_{\rm res}^{\rm tree}(p)=\mathcal{P}_0+\Delta \mathcal{P}_{\rm res}^{\rm tree}(p)$.
Moreover, the two corrections being a priori out of phase, there will necessarily be some scales at which the loop will be infinitely larger than the tree.
Of course, this does not signal an overall loss of predictivity for our EFT, and one should rather focus on the relative size of the amplitudes of the oscillations.
Furthermore, we have seen already that the two phases converge in the limit of rapid oscillations $\omega \rightarrow \infty$, which is the regime in which perturbativity could be questioned.
Focusing on this regime, we therefore simply ask that $\mathcal{P}_0 \times\rho_{\rm 1\text{-}loop}(\omega) \ll \rho_{\rm tree}(\omega) $, which amounts to asking that the strong coupling scale $\Lambda_{\rm res}$ verifies 
\begin{equation}
    \Lambda_{\rm res}^2 \gg H^2 \quad \text{where} \quad \Lambda_{\rm res} = \frac{\sqrt{8 \pi^2\epsilon_0 c_s}}{\omega} \mpl \, \implies \, \omega \ll \mathcal{P}_0^{-1/2}\, .
\end{equation}
We verify this expression in Fig.~\ref{fig:resonant} where we plot the amplitude
of both the tree-level and loop-level scale-dependent corrections for $\mathcal{P}_0 = A_s = 2.1 \times 10^{-9}$ observed at CMB scales. 
Perturbativity breaks down at $\omega \simeq 2\times10^4$ which indeed corresponds to $A_s^{-1/2}$, as expected. 
But measurements of the CMB anisotropies can only test frequencies $\omega \lesssim \mathcal{O}(10^2)$, as higher values would over-fit the data~\cite{Planck:2018jri}.
Furthermore, frequencies as high as $\omega \sim \mathcal{O}\left(10^4\right)$ would hardly impact the anisotropy spectra: indeed, these variations being extremely fast, they are washed out when integrated over the wave-number support of the CMB transfer functions.  
Our results thus imply that the frequency regime tested by CMB (and large-scale-structure) experiments is safe from strong coupling.

It is useful to compare this perturbativity bound with the scales controlling the background evolution.
For the resonant profile in Eq.~\eqref{eq:eps_res}, the fractional variation of $\epsilon$ over one Hubble time is $\dot\epsilon/(H\epsilon)=\eta\sim A_{\rm res}\omega$, so the background remains slowly varying in the usual slow-roll sense provided $A_{\rm res}\omega\ll1$. 
By contrast, the second slow-roll parameter $\eta_2=\ddot f/(\dot f H)$ scales as $\eta_2\sim\omega$ and is therefore not suppressed by $A_{\rm res}$, reflecting the fact that the oscillations may be rapid while remaining small in amplitude.

From an EFT viewpoint the relevant quantity associated with the time dependence is the physical frequency $E_{\rm var}\sim\omega H$ from Eq.~\eqref{eq:evar}.
In the Goldstone description the dominant quartic interaction induced by the resonant background scales as $\epsilon\,\eta\,\eta_2\sim\epsilon_0 A_{\rm res}\omega^2$, which explains the $\omega^2$ growth of loop corrections and ultimately leads to the perturbativity bound $\omega^2\mathcal P_0\ll1$ derived above.\footnote{Canonically normalising the interaction suggests an interaction scale $\Lambda_{\rm sc}\propto 1/(\omega\sqrt{A_{\rm res}})$ from Eq.~\eqref{eq:SCScale}. Since this scale diverges as $A_{\rm res}\to0$, it does not represent the intrinsic cut-off of the Goldstone EFT and therefore does not control the perturbativity of the observable.}

Therefore two independent requirements appear in the resonant scenario: adiabatic background evolution demands $\omega\ll \Lambda_{\rm EFT}/H$, while perturbative control of the resonant signal requires $\omega^2\mathcal P_0\ll1$. 
In the parameter range relevant for observations both conditions are comfortably satisfied.

\subsection{Sharp features}

\begin{figure}
        \centering
        \includegraphics[width=\textwidth]{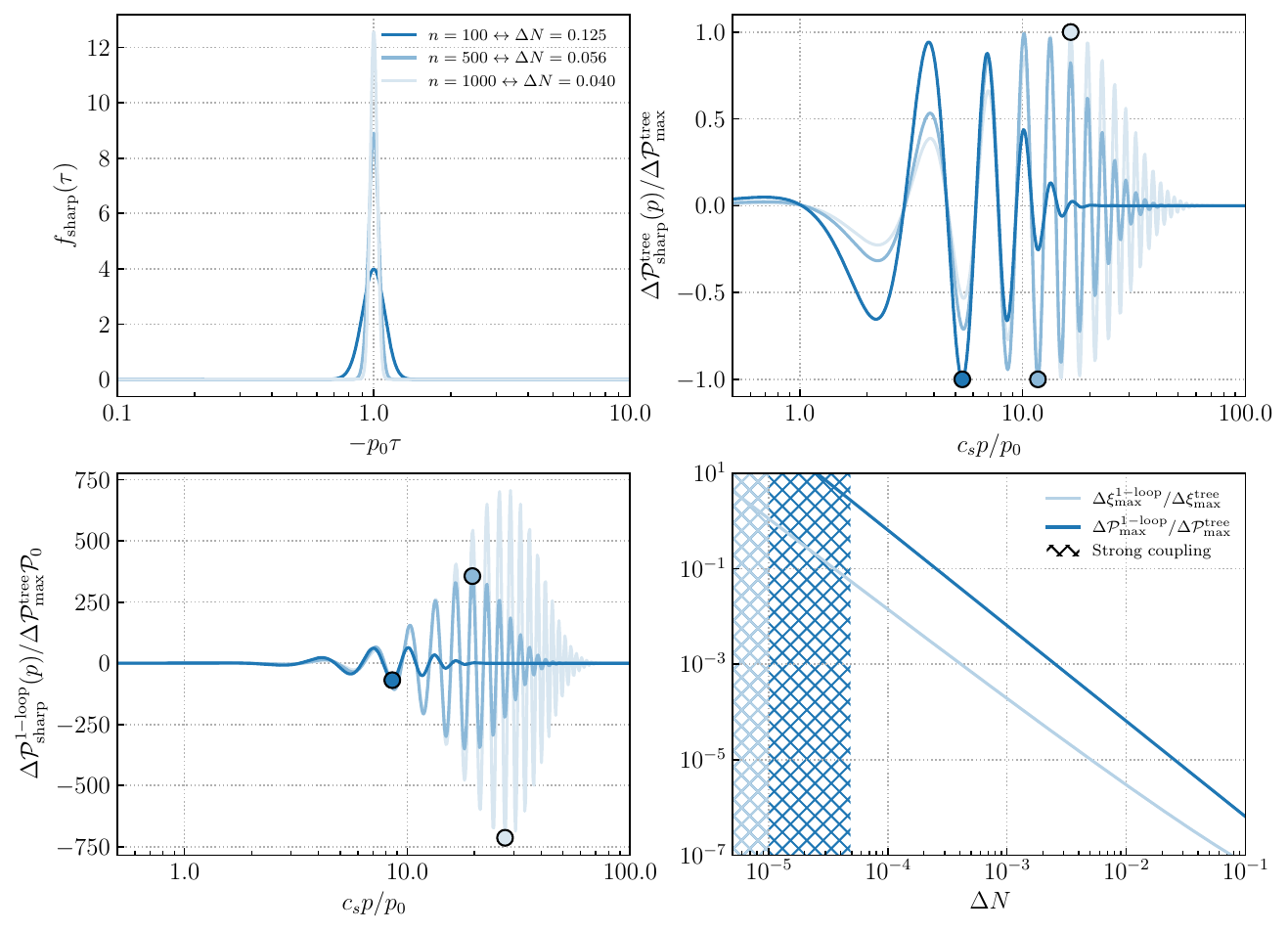}
        \caption{[Top-Left] Time-dependent correction to the first SR parameter $\epsilon$ for the sharp feature. Tree-level [Top-Right] and 1-loop [Bottom-Left] corrections to the scalar power spectrum at the end of inflation. In the top-right and bottom-left panels, we normalise the spectra by the maximum amplitude of the correction to the tree-level power spectrum. The maxima of the power spectra are marked with dots. [Bottom-Right] Ratio of the maximum amplitude of the 1-loop to tree-level spectra and real space 2-pt function at the end of inflation.  The hatched regions signal the strong coupling regimes from the two criteria.}
        \label{fig:sharp}
\end{figure}

\paragraph{Setup.}
We now turn to the case of sharp features, still modelled as Eq.~\eqref{eq:eps_time_dep}, but with 
\begin{equation}
\label{eq:eps_res}
A
=
A_{\rm sharp} \,, \quad f(\tau)=f_{\rm sharp}(\tau) \equiv 
\frac{1}{(n-1)!}\left(\frac{n\tau}{ \tau_0}\right)^n\,\exp\left(-\frac{n \tau}{ \tau_0}\right)
\,,
\end{equation}
which is normalised to $\int_{-\infty}^0\dd\ln(-\tau) \,f_{\rm sharp}(\tau)=1$.
With this choice, there is no residual symmetry in the time domain, so any function of time is now allowed for the coefficients of the EFT operators, see also~\cite{Cespedes:2012hu,Bartolo:2013exa,Cannone:2014qna,Ballesteros:2021fsp} for earlier studies of sharp features in the EFT of inflationary perturbations.
The parameter $p_0 \equiv -1/\tau_0 $ defines the scale of the sharp feature.
The sharpness of the feature is controlled by the positive integer $n$, with larger values of $n$ being associated with sharper features. In fact, to better discuss the physics of this model we can define the duration of the feature in $e$-folds $\Delta N$ as:
\begin{equation}
\label{eq:deltaN}
\int_{\tau_0 e^{\Delta N /2}}^{\tau_0 e^{-\Delta N /2}}\dd\ln(-\tau)\, f_{\rm sharp}(\tau)=C\,,
\end{equation}
where $C<1$ is an arbitrary constant, which we set to $C=1/2$.  Solving the integral in~\eqref{eq:deltaN}, and taking the large $n$ limit--which is the requirement for the feature to be sharp--we find:
\begin{equation}
    \Delta N =\sqrt{\frac{\pi}{2 n}}.
\end{equation}
We show in the upper left panel of Fig.~\ref{fig:sharp} the explicit time dependence of $f_{\rm sharp}$, for a few values of  $\Delta N$.

\paragraph{Tree-level scale dependence.}

At late times, $\tau\to 0$, the fractional tree-level scale-dependent correction takes the closed form, at leading  order in $A_{\rm sharp}$ (see App.~\ref{appA} for the details),
\begin{align}
\label{eq:result_integral_P_sharp}
\frac{\Delta\mathcal{P}_{\rm sharp}^{\rm tree}(p)}{\mathcal{P}_0}
&= A_{\rm sharp} \,\Re \left\{i\frac{n}{
  (n-1) x} \frac{ \left[1-2 i x (n+1)/n -2 x^2 (n+1)/n\right]}{\left(1 -2 i x /n\right)^{n+1}}\right\}\,,
\end{align}
where we have defined $x=c_s p /p_0$.
The amplitude of this tree-level scale-dependent correction vanishes at large scales, being $\propto x^2$ as $x\ll 1$, and at small scales, being $\propto x^{-n}$ as $x \gg 1$.

However, it is hard to read the details of the feature from this analytical expression, especially for intermediate values of $x$ where most of the information hides.
In this respect, it is interesting to notice that Eq.~\eqref{eq:result_integral_P_sharp} admits a very simple resummation.
First, we see that, for large $n$, $\left(1- 2 i x /n\right)^{-n} = \exp \left( 2 i x \right) \left[ 1 - 2 x^2/n + \mathcal{O}\left(n^{-2}\right)  \right]$, where the coefficients of order $n^{-k}$ in the expansion are always polynomials of order $2k$ in $x$.
This guides us towards defining the scaling relation $\beta \equiv x/\sqrt{n}$, under which the result reads, for large $n$, $\Delta\mathcal{P}_{\rm sharp}^{\rm tree}(p)/(\mathcal{P}_0 A_{\rm sharp})\underset{n \gg 1,\,\beta \, \text{fixed}}{=} \Re\left[-2 i \beta \sqrt{n} \exp \left(2 i \beta \sqrt{n} - 2 \beta^2 \right)\right]$.
Now that this limit has been taken, we can restore the physical $(x, n)$ dependence, yielding
\begin{equation}
\label{eq:large-n-tree-PS}
    \frac{\Delta\mathcal{P}_{\rm sharp}^{\rm tree}(p)}{\mathcal{P}_0 A_{\rm sharp}}\underset{n \gg 1, \, x^2/n \, \text{fixed}}{=}\Re \left[-2 i x\, e^{2 i x}\right]e^{-\frac{ 2x^2}{n}}\ \,.
\end{equation}
We show in Fig.~\ref{fig:large-n} in App.~\ref{appB} the impressive agreement between the full result and this approximate formula.
Of course, the formula is a priori not valid for arbitrarily small or large values of $x$.
But this is unimportant for our purpose because we already know that the feature actually vanishes in both limits.
More interestingly, we now have a better physical understanding of the feature where it has most of its support.
This includes: the explicit unveiling of the usual oscillatory pattern in $p$-space for sharp features (rather than $\ln p$-space for resonant ones), the linear growth of the envelope for $1<x<\sqrt{n}$, the Gaussian damping for $x>\sqrt{n}$, etc.
In particular, we can describe very well the maximum of the envelope, which we find to be
\begin{equation}
\label{eq:P_max_tree}
    c_s \frac{p_{\rm max}^{\rm tree}}{p_0} = \frac{\sqrt{n}}{2} \propto \Delta N^{-1} \,\, \implies \,\,\frac{\Delta\mathcal{P}_{\rm max}^{\rm tree}}{\mathcal{P}_0 A_{\rm sharp}} =   \sqrt{\frac{n}{e}}  \times \text{phase} \propto \Delta N^{-1} \,.
\end{equation}
This will be important for discussing perturbativity aspects later on.

All these features can be seen in the upper right panel of Fig.~\ref{fig:sharp}.

\paragraph{Loop-level scale dependence.}

The procedure is the same as in the resonant feature case, with the additional complication that it involves time integrals of the form
\begin{equation}
    I_n(u) \equiv  \int^u_\infty \dd v\,v^n\, e^{Av}\,{\rm Ei}(B v),
\end{equation}
where $n$ is a positive integer,  and ${\rm Re}\left(A+B\right)<0$.
One can show, by induction, the following identity:
\begin{align}
    I_n (u) =& (-1)^n A^{-n-1} \left\{{\rm Ei}(B u) \Gamma (n+1,-A u)-n! \, {\rm Ei}[(A+B)u]\right\}\notag\\&+\sum _{k=1}^n (-1)^{n-2 k} (n-k)!
     \frac{A^{k-n-1}}{(A+B)^{k}} \binom{n}{k} \Gamma [k,-(A+B) u]\,,
\end{align}
where $\Gamma$ is the incomplete Gamma function and $\rm Ei$ is the exponential integral function.
We make use of this identity to obtain the analytical results that follow.

As before, we simply quote the final result, at late times, $\tau \rightarrow 
0$, for the fractional correction at one-loop level:
\begin{align}
\label{eq:result_integral_P_sharp_loop}
\frac{\Delta\mathcal{P}_{\rm sharp}^{\rm 1\text{-}loop}(p)}{\mathcal{P}_0^2}
=&  A_{\rm sharp} \Re \left(\frac{3}{8 x} \left(1-2 i x/n\right)^{-n} \left\{  \left(1-2 i x/n\right)^{-3} p_n^{(5)}(x) \right.\right.\\
& \left. \left. - 4 i x^2 \Phi\left[\left(1-2 i x/n\right)^{-1}, 1, n\right]\right\}\right)  + \frac{\Delta \mathcal{P}_{\rm sharp}^{\rm logs} (p)}{\mathcal{P}_0^2} \nonumber \,,
\end{align}
again at leading order in $A_{\rm sharp}$, where $\Delta \mathcal{P}_{\rm sharp}^{\rm logs} (p)$ includes various logarithmic terms such as $\ln (H/\mu)$ and $\ln(\Lambda_{\rm IR}/p)$.
$\Phi[z,s,a]$ is the Lerch transcendent, and for $s=1$ and  $a=n$ a positive integer, it can be rewritten in terms of polynomials and a logarithm of $z$.
This expression, together with the one for $\Delta \mathcal{P}_{\rm sharp}^{\rm logs} (p)$ and for the complex polynomial $p_n^{(5)}(x)$ of order 5 in $x=c_s p/p_0$ can be found in App.~\ref{appB}.
The amplitude of this loop-level scale-dependent correction vanishes both at large scales, being $\propto x^2$ as $x\ll 1$, and at small ones, being $\propto  x^{1-n}$ as $ x \gg 1$.

Interestingly, we can also perform a large-$n$ resummation of the finite pieces of the loop correction.
Indeed, we note that the same scaling with $\beta = x/\sqrt{n}$ holds.
Our task here is complicated by the $\Phi$ function, so here we will only quote the large-$n$ limit of the loop corrections without it.
In App.~\ref{appB}, see Fig.~\ref{fig:large-n}, we prove that this approximation is indeed an excellent one.
Following the same steps as in the tree-level case, restoring $(x,n)$ dependence, we have:
\begin{equation}
\label{eq:large-n-loop-PS}
    \frac{\Delta\mathcal{P}_{\rm sharp}^{\rm 1 \text{-} loop, \,finite}(p)}{\mathcal{P}_0^2 A_{\rm sharp}}\underset{n \gg 1, \,  x^2/n \, \text{fixed}}{=}
    \Re \left[-3i x^3\, e^{2 i x}\right]e^{-\frac{ 2x^2}{n}} \,.
\end{equation}
It is striking to realise that such a simple expression represents the scale dependence from renormalised loop corrections.
We have indeed successfully recovered all its properties around the peak: $\cos(2 x)$ oscillations, cubic growth for $1<x < \sqrt{n}$, Gaussian damping for $x>\sqrt{n}$, etc.
Importantly, this formula also allows us to describe the maximum of the envelope,
\begin{equation}
\label{eq:P_max_loop}
    c_s \frac{p_{\rm max}^{\rm 1\text{-} loop}}{p_0} = \frac{\sqrt{3n}}{2} \propto \Delta N^{-1} \,\, \implies  \,\, \frac{\Delta\mathcal{P}_{\rm max}^{\rm 1 \text{-} loop}}{\mathcal{P}_0^2 A_{\rm sharp}} =   \frac{9}{8}\sqrt{\frac{3n^3}{e^3}}  \times \text{phase} \propto \Delta N^{-3}\,,
\end{equation}
a key information to discuss perturbativity indeed.

We represent the exact result, Eq.~\eqref{eq:result_integral_P_sharp_loop}, for the loop-level scale-dependent correction, in the lower left panel of Fig.~\ref{fig:sharp}.
Comparing to the upper right panel, one can already see by eye that the most boosted wavelengths correspond to smaller scales (larger $p$) at loop level than at tree level, as further indicated by our analytical formula $p_{\rm max}^{\rm 1\text{-} loop}/p_{\rm max}^{\rm tree} = \sqrt{3}$. 
Moreover, the amplitude of the feature grows faster in the loop than in the tree when the sharpness of the feature increases, i.e. when $\Delta N$ decreases, again as we see from the analytical formulas above.
Finally, we also recover numerically what the analytical expressions have already taught us: the amplitude of the fully renormalised scale-dependent one-loop correction vanishes both at very large scales and very small ones, just like tree-level corrections, namely,
\begin{align}
    \frac{\Delta\mathcal{P}_{\rm sharp}^{\rm 1\text{-}loop}(p)}{\mathcal{P}_0^2 A_{\rm sharp}} &\underset{p\ll p_0}{\propto} \left(\frac{p}{p_0}\right)^{2} \,, \\
    \frac{\Delta\mathcal{P}_{\rm sharp}^{\rm 1\text{-}loop}(p)}{\mathcal{P}_0^2  A_{\rm sharp}} &\underset{p\gg p_0}{\propto} 
        \left(\frac{p}{p_0}\right)^{1-n}  \,.
\end{align}
This is an important result in light of the recent debates  about whether loop corrections from gravitational non-linearities, i.e. the ones that we have taken into account, could efficiently transfer an excess of power from small cosmological scales to CMB ones~\cite{Cheng:2021lif,Inomata:2022yte,Kristiano:2022maq,Riotto:2023hoz,Firouzjahi:2023aum,Choudhury:2023vuj,Motohashi:2023syh,Franciolini:2023lgy,Tasinato:2023ukp,Cheng:2023ikq,Fumagalli:2023hpa,Maity:2023qzw,Davies:2023hhn,Tada:2023rgp,Iacconi:2023ggt,Inomata:2024lud,Firouzjahi:2024psd,Caravano:2024tlp,Ballesteros:2024zdp,Kristiano:2024vst,Kawaguchi:2024rsv,Fumagalli:2024jzz,Caravano:2024moy,Ruiz:2024weh,Firouzjahi:2024sce,Sheikhahmadi:2024peu,Inomata:2025bqw,Fang:2025vhi,Firouzjahi:2025gja,Firouzjahi:2025ihn,Inomata:2025pqa,Iacconi:2026uzo,Li:2026vrn}.
Of course, the particular scenario that we have investigated, a sharp feature with a small amplitude, is physically distinct from the transient ultra-slow-roll scenario considered in these works.
However, we claim that our result is the first renormalised one-loop correction explicitly computed in a scale-dependent scenario with a localised feature along the inflationary dynamics.
It is quite striking that the vanishing of the loop correction at small and large $p$ is valid irrespective of the details of the sharp feature, and in particular of the value of $\Delta N$. In particular, although the support of the feature does extend to larger $p\gg p_0$ as $\Delta N$ decreases, this has no impact on the power spectrum for scales that are super-horizon at the time of the feature for which $p\ll p_0$ and the loop correction goes to zero as $(p/p_0)^2$, i.e. independently of $\Delta N$. 
In this regard, we would like to emphasise that although the renormalised spectrum is finite, several individual contributions to it diverge in the limits of large and small momenta. For example, the bare loop correction induced by the spatial and time derivatives in the quartic interactions behaves as $\pm\,A_{\rm sharp}  \mathcal{P}_0^2 \pi (c_s p/p_0)/2$ respectively in the limit $p\gg p_0$, so that their sum vanishes. Furthermore, the bare loop from spatial derivatives behaves in the limit $p\ll p_0$ as $A_{\rm sharp} \mathcal{P}_0^2 \pi (c_s p/p_0)^{-1} n /2(n-1)$. This divergence is cancelled by a term with an opposite sign arising from the finite part of the two counter-terms in Eq.~\eqref{eq:solved_counter-terms}.
We believe that this result might guide the resolution of the aforementioned debates about loop corrections for scenarios with interesting small-scale phenomenology, such as a large production of primordial black holes and scalar-induced gravitational waves.

\paragraph{Perturbativity.}
For this case of a sharp feature, the question of perturbativity is harder to address.
We propose two ways to define perturbativity and we compare them.

First, we ask that the loop correction to the power spectrum at its maximum be smaller than the tree-level correction at its own maximum, namely
\begin{equation}
    \Delta\mathcal{P}_{\rm max}^{\rm 1\text{-}loop} \ll \Delta\mathcal{P}_{\rm max}^{\rm tree}\,.
\end{equation}
This criterion was also used in Ref.~\cite{Ballesteros:2024zdp} in the context of a different model featuring an intermediate USR stage.
Making use of the formulas derived for $x \sim \sqrt{n}$, we find the following condition on the strong coupling scale:
\begin{equation}
\Lambda_{\rm sharp}^2  \gg H^2  \quad \text{where} \quad \Lambda_{\rm sharp} \equiv \Delta N \sqrt{8 \pi^2 \epsilon_0 c_s} \mpl \, \implies \, 
    \Delta N \gg \mathcal{P}_0^{1/2} \,,
\end{equation}
which for $\mathcal{P}_0=A_s =2.1 \times 10^{-9}$ is $\Delta N \gg 5 \times 10^{-5}$.

Second, we ask that the correction to the real-space two-point correlation function from the primordial feature be smaller at the loop level than at the tree one~\footnote{Note that we have not done that for the resonant feature, because in that case the real-space two-point correlation function is again an oscillating function, this time in $\ln r$-space, and nothing is gained compared to the Fourier space analysis.}.
Namely, we define:
\begin{align}
 \Delta   \xi(r) &=  \Delta\xi_{\rm tree}(r) + \Delta\xi_{\rm 1\text{-} loop}(r) \,, \quad \text{with} \\
     \Delta\xi_{\rm i}(r) &= \int \frac{\dd^3 \vec{p}}{(2\pi)^3}\, e^{i \vec{p} \cdot \vec{r}}\,\Delta P_{\rm sharp}^{\rm i}(p)  \,=\,\int \frac{\dd  p}{p} \frac{\sin (p r)}{pr}   \Delta \mathcal{P}_{\rm sharp}^{\rm i}(p) \nonumber \,,
\end{align}
where ${\rm i} = {\rm tree},\,{\rm loop}$. In the second equality, we performed the angular integrals explicitly, and switched from the dimensionful to the dimensionless power spectrum, and isotropy is responsible for the $r \equiv |\vec{r}|$-dependence only.
We can again make use of the analytical formulas for $x\sim\sqrt{n}$ and $n\gg 1$ to find:
\begin{align}
\label{eq:large-n-real-space}
   \frac{\Delta\xi_{\rm tree}(r)}{\mathcal{P}_0 A_{\rm sharp}} &= \frac{n}{8 u}\left\{(u+2)^2 \,   {}_2F_2\left[1,1;\frac{3}{2},2;-\frac{n}{8}(u+2)^2\right] - \{ (u+ 2) \rightarrow (u-2) \} \right\} \,, \nonumber \\
   \frac{\Delta\xi_{\rm 1\text{-} loop}(r)}{\mathcal{P}_0^2 A_{\rm sharp}} & = \frac{3}{8 u} \sqrt{\frac{n^3}{2}} \left\{(u+2)D\left[\sqrt{\frac{n}{8}} (u+2) \right] - \{ (u+ 2) \rightarrow (u-2) \} \right\}\,,
\end{align}
with $u= p_0 r /c_s$, ${}_2F_2$ is the generalised hypergeometric function and $D$ is the Dawson function denoted as \texttt{DawsonF} in Mathematica.
Thanks to these simple expressions, we can read that the maximum of $\Delta \xi_{\rm i}(r)$ is always located at $u_{\rm max} = 2 \iff r_{\rm max}=2 c_s /p_0$, which gives the unambiguous perturbativity criteria:
\begin{equation}
     \Delta \xi_{\rm max}^{\rm 1\text{-}loop} \ll \Delta \xi_{\rm max}^{\rm tree}\,.
\end{equation}
Moreover, in the large $n$ limit, those maxima behave as:
\begin{equation}
\label{eq:xi_max}
    \frac{\Delta \xi_{\rm max}^{\rm tree}}{\mathcal{P}_0 A_{\rm sharp}} \underset{n\gg 1}{=} \frac{\ln  (8 n) + \gamma_E}{4} \quad \text{and} \quad \frac{\Delta \xi_{\rm max}^{\rm 1\text{-}loop}}{\mathcal{P}_0^2 A_{\rm sharp}} \underset{n\gg 1}{=} \frac{3 n}{16}\,.
\end{equation}
In App.~\ref{appB}, we confirm agreement between the results for the real-space two-point function in the large $n$ limit and the numerical results obtained using the full expression of the tree and 1-loop corrections to the power spectra.

Given these results, we read that perturbativity holds as long as
\begin{equation}
    n \ll 2 \times 10^{10} \,\, \implies \,\, \Delta N \gg 10^{-5} \,,
\end{equation}
where we used $\mathcal{P}_0 = A_s = 2.1 \times 10^{-9}$. 
Interestingly, the perturbativity criterion in real space is compatible with the one in Fourier space, though being slightly less restrictive.
This can be simply understood from the fact that the real-space two-point function corresponds to the integration of the Fourier-space one over many oscillations, thus decreasing the significance of the primordial feature.
We represent these two criteria for strong coupling in the lower right panel of Fig.~\ref{fig:sharp}.

It is worth stressing that the strong-coupling scale inferred from the
canonically normalised quartic operator in Eq.~(\ref{eq:SCScale}) does
not by itself control perturbativity in the sharp case. Instead, the
breakdown of perturbation theory is determined by the ratio between the
loop and tree-level feature contributions, which leads to the bound
$\Delta N \gg \sqrt{\mathcal P_0}$ obtained above. This condition also
ensures that the background evolution remains adiabatic in the EFT
sense. A feature of duration $\Delta N$ introduces a characteristic
variation scale $E_{\rm var}\sim H/\Delta N$ for the time-dependent
couplings, so that the bound above implies $E_{\rm var}\ll\Lambda_{\rm EFT}$,
consistent with the validity condition discussed around
Eq.~(\ref{eq:evar}).

\section{Conclusion}
The near scale invariance of inflationary correlation functions, strongly supported by observations, together with the technical difficulty of loop calculations, has confined most studies of loop corrections to scale-invariant backgrounds, namely de Sitter spacetime. Recently, however, loop effects in strongly time-dependent backgrounds have attracted renewed attention, particularly in scenarios where large violations of slow roll may lead to abundant primordial black hole production~\cite{Cheng:2021lif,Inomata:2022yte,Kristiano:2022maq,Riotto:2023hoz,Firouzjahi:2023aum,Choudhury:2023vuj,Motohashi:2023syh,Franciolini:2023lgy,Tasinato:2023ukp,Cheng:2023ikq,Fumagalli:2023hpa,Maity:2023qzw,Davies:2023hhn,Iacconi:2023ggt,Inomata:2024lud,Firouzjahi:2024psd,Caravano:2024tlp,Ballesteros:2024zdp,Kristiano:2024vst,Kawaguchi:2024rsv,Fumagalli:2024jzz,Caravano:2024moy,Ruiz:2024weh,Firouzjahi:2024sce,Sheikhahmadi:2024peu,Inomata:2025bqw,Fang:2025vhi,Firouzjahi:2025gja,Firouzjahi:2025ihn,Inomata:2025pqa,Iacconi:2026uzo,Li:2026vrn}. These situations fall outside the scale-invariant framework and raise new challenges for the computation and renormalisation of loop corrections.

In this work we develop a general procedure to renormalise the one-loop primordial power spectrum of curvature perturbations in inflationary backgrounds that depart from exact de Sitter symmetry. The analysis is performed within the EFT of inflationary perturbations, which allows one to treat controlled departures from scale invariance arising from time-dependent backgrounds. We work in the decoupling limit ($\epsilon\ll1$) and assume a hierarchy among slow-roll parameters such that $\eta\ll\eta_2$. 
Under these conditions, the structure of ultraviolet divergences and tadpoles can be determined in full generality and cancelled by a finite set of local counter-terms consistent with the EFT symmetries.
An important consequence of this analysis is that the associated counter-terms can be identified without specifying the detailed time dependence of the background expansion, assuming only that the mode functions can be adiabatically connected
to the Bunch–Davies vacuum at early times.
In particular, renormalisation requires counter-terms generated by EFT operators that are degenerate with other unitary-gauge building blocks at tree level, such as $\mathcal{L}\supset-\tilde{M}_1^3(\tau)\delta K$. The time dependence of the background and the structure of loop contractions break this degeneracy at one loop and make such operators necessary for renormalising the theory. This contrasts with the standard scale-invariant case, where such operators are not required for the renormalisation of the scalar power spectrum~\cite{Braglia:2025cee,Braglia:2025qrb}.

To illustrate these general results, we apply the formalism to two explicit examples of violations of scale invariance: resonant and sharp features. In the regime where the modification of the background is small, the in-in integrals can be evaluated analytically while keeping the mode functions equal to their de Sitter form. The two scenarios lead to qualitatively different loop corrections.
For resonant features, the scale dependence of the renormalised power spectrum at one loop coincides with the tree-level result, so the loop correction simply rescales the amplitude of the oscillatory signal. The result can nevertheless be used to set a perturbativity bound $\omega\ll A_s^{-1/2}$ on the dimensionless frequency of the feature.
For sharp features, the situation is different: the one-loop correction modifies the envelope of the oscillatory spectrum, shifting the maximum to higher wave-numbers and leading to the perturbativity bound $\Delta N\gg10^{-5}$ on the sharpness of the feature.

The renormalised result contains infrared (but no secular) logarithmic divergences that deserve further investigation.
A logarithmic dependence on the renormalisation scale $\mu$ also appears, as expected in dimensional regularisation.
Fixing the renormalisation condition $\mu\propto H$ removes this contribution from observables.
For sharp features, the remaining loop correction does modify the scale dependence of the power spectrum. However, observing such a scale dependence would not necessarily constitute a direct detection of loop effects. Local contributions generated by loops can be absorbed into a redefinition of background quantities such as the slow-roll parameters or, more generally, the
Wilson functions of the EFT. A suitable time dependence of these functions can therefore reproduce the one-loop spectrum already at tree level, and this procedure can be repeated order by order in perturbation theory, similarly to the construction of a renormalised potential in quantum field theory. Genuine loop signatures must therefore arise from observables that are not degenerate with such local redefinitions.
This may occur in higher-order correlation functions such as the bispectrum~\cite{Bhowmick:2024kld}, as recently pointed out in~\cite{Ballesteros:2025nhz}, or in observables probing different sectors of the theory.
For instance, the tensor power spectrum is scale invariant at tree level in these models but may acquire scale dependence at loop level, potentially modifying the usual consistency relations.

Beyond their theoretical significance, these results are also relevant for phenomenological models of primordial features that provide good fits to CMB residuals and satisfy the assumptions of our analysis. In this region of parameter space the models remain perturbatively consistent.

Finally, we comment on scenarios in which violations of slow roll are sufficiently strong to significantly amplify scalar perturbations, as often considered in the recent literature. In such situations, the hierarchy $\eta\ll\eta_2$ may break down, so additional diagrams contribute, and the perturbative treatment adopted here may no longer apply. Nevertheless, one result of our analysis is directly relevant: the renormalised one-loop correction induced by a sharp feature vanishes for modes that cross the horizon well before or well after the time of the feature. Several works working within different setups and approximations~\cite{Kristiano:2022maq,Fumagalli:2023hpa,Fumagalli:2024jzz,Kawaguchi:2024rsv,Tada:2023rgp,Inomata:2024lud} have argued that the loop correction generated by a feature at time $\tau_0=-p_0^{-1}$ can be significantly amplified at scales $p\ll p_0$. We show that individual contributions can grow in this limit but cancel once all diagrams are consistently combined and the theory is renormalised. Our results clarify this issue and provide a first step toward a systematic treatment of loop corrections in strongly time-dependent inflationary backgrounds.

\acknowledgments
We thank Jason Kristiano, Vincent Vennin and Zhong-Zhi Xianyu for useful discussions. We also thank the anonymous Referee for useful comments that improved the clarity of our paper. MB an SC thank LPENS for hosting them while this work was in progress.
MB has received funding from the European Union’s Horizon 2020
research and innovation programme under the Marie Skłodowska-Curie grant
agreement No 101205460.
SC is supported in part by the STFC Consolidated Grants ST/T000791/1 and ST/X000575/1
and by a Simons Investigator award 690508.

\appendix

\section{UV divergences in dimensional regularisation}
\label{appC}

In this Appendix, we justify that the divergent part of the loop diagram is
given by~\eqref{eq:UV} provided the short--distance modes are in the
Bunch--Davies (adiabatic) vacuum~\cite{ParkerFulling1974,BirrellDavies,AndersonParker1987} (See eg. \cite{Kristiano:2025ajj,Ballesteros:2025nhz} for a recent application of these techniques within the context of inflation). Throughout, we use dimensional
regularisation for the momentum integral in $d=3+\delta$ spatial
dimensions. In this scheme only logarithmic UV divergences generate a
$1/\delta$ pole; pure power divergences are absent in minimal
subtraction. 

In the interaction picture, the mode functions appearing in the in-in integrals are the ones of the free theory, which in our case is set by the bare quadratic EFT action for the Goldstone mode.
We define the canonically normalised mode functions in conformal time as
\begin{equation}
    v_k(\tau)\equiv z(\tau)\,\pi_k(\tau)\,, \quad  z^2(\tau)\equiv 2a^2(\tau)\frac{\epsilon(\tau)M_{\rm Pl}^2}{c_s^2(\tau)}\,;
\end{equation}
they verify the standard Sasaki-Mukhanov equations (here and in the following we drop the explicit time dependence of the background quantities),
\begin{equation}
    v_k''+\omega_k^2\,v_k=0\,,
\qquad
\omega_k^2=c_s^2 k^2-\frac{z''}{z} \,.
\label{eq:MuSAs}
\end{equation}

The UV regime is defined as the domain where the equation
admits an adiabatic solution. 
To look for this adiabatic solution, we substitute the usual WKB ansatz 
\begin{equation}
v_k(\tau)=\sum_{\pm}\frac{\alpha^\pm}{\sqrt{2W_k(\tau)}}
\exp\!\Big(\pm i\int^\tau W_k(\tau')\,d\tau'\Big)
\end{equation}
into~\eqref{eq:MuSAs}, and we find that $W_k(\tau)$ must verify the following non-linear second-order differential equation:
\begin{equation}
    W_k^2=\omega_k^2-\frac{1}{2}\frac{W_k^{\prime\prime}}{W_k} + 3\frac{W_k^{\prime 2}}{W_k^2}  \,.
    \label{eq:sol_Wk_generic}
\end{equation}
Moreover, imposing that the solution matches the Bunch-Davies vacuum in the $k \rightarrow \infty$ limit where $\omega_k = {\rm const} = k \implies W_k = k$, we find
\begin{equation}
    \alpha^+ = 0 \,, \quad \alpha^- = 1 \,.
\end{equation}
So far, the discussion has been generic and no approximation has been used.

\subsection{Leading-order WKB approximation}

We first truncate Eq.~\eqref{eq:sol_Wk_generic} at leading order in the background time variation, which gives 
\begin{equation}
    W_k = \omega_k
\end{equation}
Substituting this back into Eq.~\eqref{eq:sol_Wk_generic} we get the usual self-consistent conditions for the WKB approximation to hold:
\begin{align}
\left|\frac{\omega_k'}{\omega_k^2}\right|^2\ll1\,,
\qquad
\left|\frac{\omega_k''}{\omega_k^3}\right|\ll1\,,
\label{eq:adiab_cond}
\end{align}
which require that $\omega_k$ varies slowly compared with the oscillation
time scale $1/\omega_k$. 
Replacing the definition of $\omega_k$, a
sufficient set of inequalities is
\begin{align}
(c_s k)^6 &\gg \left|\frac{\dd}{\dd \tau}\left(\frac{z''}{z}\right)\right|^2\,,
\quad
c_s^2 k^2 \gg \left|\frac{c_s'}{c_s}\right|^2\,, \\
(c_s k)^4 &\gg \left|\frac{\dd^2}{\dd \tau^2}\left(\frac{z''}{z}\right)\right|\,,
\quad
c_s^2 k^2 \gg \left|\frac{c_s^{\prime\prime}}{c_s}\right|\nonumber \,.
\label{eq:UVconds}
\end{align}

The leading WKB approximation fixes uniquely the dominant contributions to the large-$k$ asymptotic expansion of the
equal-time correlator. 
Indeed, its general expression gives
\begin{equation}
|v_k(\tau)|^2=\frac{1}{2\omega_k(\tau)} \,, 
\qquad
|\pi_k(\tau)|^2=\frac{|v_k(\tau)|^2}{z^2}.
\end{equation}
However, in dimensional regularisation the $1/\delta$ pole comes from the
first term in the large-$k$ expansion that behaves as $k^{-3}$ at fixed
$\tau$, since
\begin{equation}
\int \dd^{3+\delta}k\;k^{-3}\ \propto\ \int^\infty \dd k\,k^{-1+\delta}\ \sim\ \frac{1}{\delta}.
\label{eq:logcriterion_app}
\end{equation}
But the dominant piece from $\omega_k = k$ gives $|v_k|^2\propto 1/k$ and instead produces only a power
divergence (which is absent in minimal subtraction).
To isolate the pole, one therefore consistently expands $\omega_k$ for large $k$,
\begin{equation}
\label{eq:omegak-largek}
\omega_k(\tau)
= c_s k\left[1-\frac{1}{2 c_s^2 k^2}\frac{z''}{z}
+\mathcal{O}(k^{-4})\right]\,,
\end{equation}
and hence
\begin{align}
|v_k(\tau)|^2
&=
\underbrace{\frac{1}{2c_s}\frac{1}{k}}_{\text{power div. (no }1/\delta\text{ in MS)}}
\;+\;
\underbrace{\frac{1}{4 c_s^3}\frac{z''}{z}\,\frac{1}{k^3}}_{\text{log div. }(\propto 1/\delta)}
\;+\;\mathcal{O}(k^{-5})\,,
\label{eq:piUV_expand}
\end{align}
and similarly for $|\pi_k(\tau)|^2$.
Equation~\eqref{eq:piUV_expand} makes explicit that the coefficient of the $1/\delta$
pole is controlled by the universal $k^{-3}$ tail fixed by the adiabatic (WKB) expansion.
The leading $k^{-1}$ behaviour is also fixed but it corresponds to a power law divergence absent in dimensional regularisation.
Not only does this prove that the contractions kept in Eq.~\eqref{eq:UV} indeed contain UV divergences that must be cancelled (for all external times and all external wave-numbers), but also that these divergences are universal.
This neatly explains why we were able to cancel them using a finite set of local counter-terms consistent with the EFT symmetries and without having to assume a particular background evolution.

In order to prove that the other contractions \textit{do not} result in additional UV divergences in the regime that we are considering, we need to go to next-to-leading order in the WBK approximation.
For example, spatial derivative contractions like $c_s^2k^2 |v_k(\tau)|^2$ require investigation of the $\mathcal{O}(k^{-5})$ terms in Eq.~\eqref{eq:piUV_expand}.
But those can be sourced both by (i) $\mathcal{O}(k^{-4})$ corrections to the leading piece of $\omega_k$, see~\eqref{eq:omegak-largek}; (ii) corrections to $W_k=\omega_k$ from derivatives of $\omega_k$, see~\eqref{eq:sol_Wk_generic}.
Without loss of generality, but for technical clarity, we will assume a constant speed of sound in the following.

\subsection{Next-to-leading-order WKB approximation}

By solving perturbatively Eq.~\eqref{eq:sol_Wk_generic}, we find the next-to-leading-order expression
\begin{equation}
    W_k^2 = \omega_k^2  -\frac{1}{2}\frac{\omega_k^{\prime\prime}}{\omega_k} + 3\frac{\omega_k^{\prime 2}}{\omega_k^2} \,.
\end{equation}
By plugging this into Eq.~\eqref{eq:sol_Wk_generic}, we get additional self-consistent inequalities bounding the size of higher-time-derivatives compared to $c_s k$, which are not worth reporting here.

Rather, we straight write the large-$k$ asymptotic expression with further precision,
\begin{align}
    c_s^2 k^2 |\pi_k(\tau)|^2=
\underbrace{\frac{1}{2 z^2}c_s k}_{\text{power div.}}
\;+\;
\underbrace{\frac{1}{4}\frac{z''}{z^3}\,\frac{1}{c_s k}}_{\text{power div.}}
\;+\;
\underbrace{\frac{1}{z^2}\left[3\left(\frac{z''}{z}\right)^2-\frac{\dd^2 }{\dd \tau^2}\left(\frac{z''}{z}\right)\right]\frac{1}{c_s^3 k^3}}_{\text{log div.?}} + \mathcal{O}\left(k^{-5}\right)
\,.
\label{eq:piUV_expand_NLO}
\end{align}
Naively, it would seem that there could exist additional logarithmic divergences after integrating over $\dd^3 \vec{k}$.
However, the UV regime that we are probing by expanding at large $k$ is only valid provided $(c_s k \tau)^2 \gg 1$, as can be checked just by dimensional analysis of the first two terms in the above expansion.
At such early times, it is reasonable to expect that the background spacetime resembles de Sitter, an assumption which is perfectly compatible with being agnostic about the remaining of the inflationary expansion history.
With this observation, it becomes immediate to see that these possible logarithmic divergences actually vanish, indeed one has
\begin{equation}
    3\left(\frac{z''}{z}\right)^2-\frac{\dd^2 }{\dd \tau^2}\left(\frac{z''}{z}\right) = \frac{9}{16} (a H)^4 \left( 2 \epsilon + \eta +\ldots \right) \underset{\epsilon,\, \eta \, \rightarrow  \, 0}{\longrightarrow} 0 \,,
\end{equation}
with dots containing terms at higher order in the slow-roll parameters.

After straightforward but tedious manipulations, it can be seen that the same cancellations happen for the $k^{-3}$ terms inside the other contractions involving derivatives of $\pi_k$, namely
\begin{align}
    \pi^\prime_k \pi_k^*+\pi^{\prime *}_k \pi_k &\underset{\epsilon,\, \eta \, \rightarrow  \, 0}{\longrightarrow} -\frac{1}{z^2}\frac{aH}{c_sk} +\mathcal{O}(k^{-5}) \,, \\
    |\pi^{\prime}_k|^2  &\underset{\epsilon,\, \eta \, \rightarrow  \, 0}{\longrightarrow} \frac{1}{2 z^2}\frac{1}{c_sk} +\mathcal{O}(k^{-5}) \,.
\end{align}
By plugging this into~\eqref{eq:bare}, we see that the only UV-divergent pieces come from the contributions $\propto |\pi_k(\tau)|^2$ without derivatives acting on it, i.e. the ones reported in Eq.~\eqref{eq:UV}.
We insist that this proof only relies on the existence of an adiabatic (WKB) regime in the UV, that is continuously related to the Bunch-Davies state.

We also note that, although the loop momentum integral is analytically
continued to $d=3+\delta$ spatial dimensions, the mode equation in this
Appendix has been written in three dimensions. This is sufficient for the
present purpose, since the coefficient of the $1/\delta$ pole is determined
by the large-$k$ asymptotics of the modes, which is unchanged by the
analytic continuation of the spatial dimension at the order relevant for
the ultraviolet divergence. The continuation $d\to3+\delta$ therefore
affects only the phase-space measure $\dd^{3+\delta}k$ in the loop
integral. Any correction to the mode equation arising away from $d=3$
would modify only finite terms, of order $\delta^0$, and not the
coefficient of the $1/\delta$ pole.
\section{Tree-level scale-dependent corrections}
\label{appA}

\begin{figure}
        \centering
        \includegraphics[width=.48\textwidth]{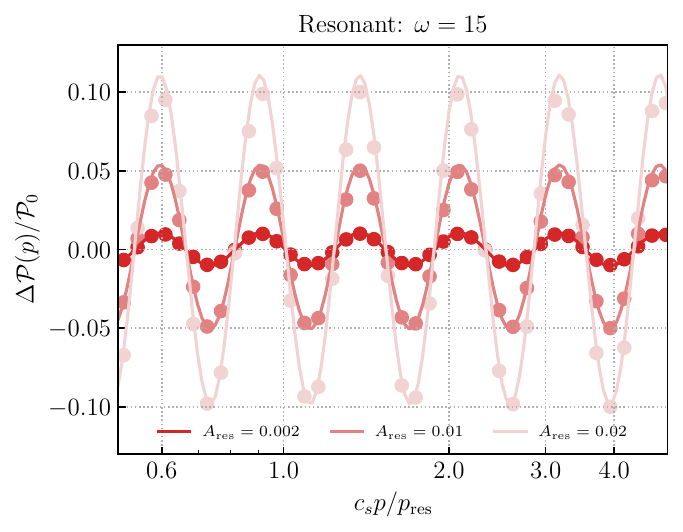}
        \includegraphics[width=.495\textwidth]{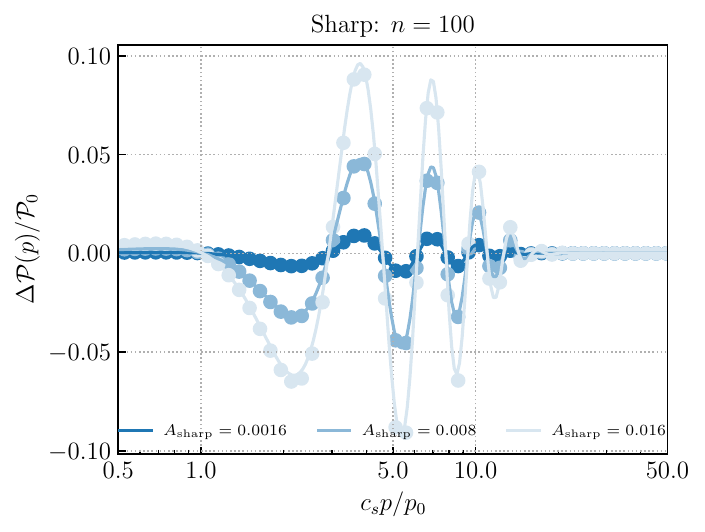}
        \caption{Comparison of the numerical results obtained by solving the EOM of $\pi$ (dots) and the analytical solution from the in-in perturbation theory at first order in the feature amplitude parameter $A$ (solid lines). We plot one illustrative parameter choice for resonant and sharp features in the left and right panels respectively, and vary the overall feature amplitude. As can be seen, the matching is perfect for tree-level corrections to the power spectrum of order $\Delta \mathcal{P}/\mathcal{P}_0\lesssim\mathcal{O}(0.1)$. }
        \label{fig:Pk_tree_res}
\end{figure}

In this Appendix, we outline our calculation of tree-level power spectrum for resonant and sharp features within the in-in formalism. We start from the second order Hamiltonian, which is given by
\begin{equation}
\label{eq:Hamiltonian_free_nonpert}
\mathcal{H}^{(2)}=\frac{c_s^2\,p_\pi^2}{4 a^3 \mpl^2H^2\epsilon} +a^3\mpl^2H^2\epsilon \frac{(\partial\pi)^2}{a^2}.
\end{equation}
In the main text, we wrote the first SR parameter as $\epsilon=\epsilon_0+\Delta\epsilon(t)$, and assumed $\Delta\epsilon(t)\ll\epsilon_0$. With this assumption, we can split the quadratic Hamiltonian into a free and an interacting Hamiltonian as
\begin{equation}
\mathcal{H}^{(2)}=
\mathcal{H}^{(2)}_{\rm free}+
\mathcal{H}^{(2)}_{\rm int},
\end{equation}
where
\begin{equation}
\label{eq:H2_free_app}
\mathcal{H}^{(2)}_{\rm free}=\frac{c_s^2\,p_\pi^2}{4 a^3 \mpl^2H^2\epsilon_0} +a^3\mpl^2H^2\epsilon _0\frac{(\partial\pi)^2}{a^2}=\frac{a^3\mpl^2H^2\epsilon_0}{c_s^2} \left(\dot{\pi}^2+c_s^2\frac{(\partial\pi)^2}{a^2}\right),
\end{equation}
and
\begin{equation}
\label{eq:H2_int_app}
\mathcal{H}^{(2)}_{\rm int}=\frac{a^3\mpl^2H^2\epsilon_0}{c_s^2}\left(\frac{\Delta\epsilon}{\epsilon_0}\right) \left(-\dot{\pi}^2+c_s^2\frac{(\partial\pi)^2}{a^2}\right)+\mathcal{O}^{n\geq2}\left(\frac{\Delta\epsilon}{\epsilon_0}\right),
\end{equation}
and we have used the relation $p_\pi=2 a^3 H^2\epsilon_0  \mpl^2 \dot{\pi} / c_s^2$, which holds for the free theory defined by the free Hamiltonian in~\eqref{eq:H2_free_app}. We can therefore proceed to compute the correction to the tree-level power spectrum perturbatively in $\Delta\epsilon(t)\ll\epsilon_0$, keeping only the first order. The result is given by:
\begin{equation}       \Delta\mathcal{P}^{\rm tree}_\pi(p,\,\tau)=\frac{p^3}{2\pi^2} 4 {\rm Im}\, \pi_p^{*2}(\tau)\, \int_{-\infty^+}^\tau\dd\tau_1\,\frac{H^2\mpl^2}{c_s^2} a^2(\tau_1)\,\Delta\epsilon(\tau_1)\, \left({\pi'}^2_p(\tau_1)-p^2\pi^2_p(\tau_1)\right),
\end{equation}
which, assuming the functional forms for $\Delta\epsilon(t)$ reported in the main text, can be solved analytically leading to Eqs.~\eqref{eq:result_integral_P_res} and~\eqref{eq:result_integral_P_sharp}.

As a cross-check, we can compare the in-in spectra to the numerical solution to the EOM of $\pi$ derived from the Hamiltonian~\eqref{eq:Hamiltonian_free_nonpert}, where the full evolution of $\epsilon(t)$ is taken into account non-perturbatively:
\begin{equation}
    \frac{\dd}{\dd t} \left[\frac{ a^3 H^2\epsilon }{  c_s^2}\dot{\pi}\right] =  a \epsilon H^2  \partial^2 \pi.
\end{equation} The in-in and the EOM formalisms were shown to be equivalent  at tree level in \cite{Chen:2015dga}. We compare the two approaches in Fig.~\ref{fig:Pk_tree_res}, where we show one illustrative choice of parameters for each model. As can be seen, the first order in-in correction matches perfectly the results obtained with the EOM approach for amplitudes of the corrections of order  $\Delta \mathcal{P}/\mathcal{P}_0\sim\mathcal{O}(0.1)$. This is therefore the order of validity of the perturbation theory developed in the main text.

\section{Explicit loop-level results}
\label{appB}

We collect all exact expressions at loop level in this appendix.

\subsection{Resonant features}

The one-loop correction from resonant features is given by Eq.~\eqref{eq:result_integral_P_res_loop}, with
\begin{align}
    \frac{\Delta \mathcal{P}_{\rm res}^{\rm logs} (p)}{\mathcal{P}_0^2 A_{\rm res}}  & = \rho_{\rm logs}(\omega) \cos \left[ \omega \ln \left( \frac{2 c_s p}{p_{\rm res}} \right) + \phi_{\rm logs}(\omega)\right] \nonumber \\
    & \quad \times \left[\gamma_E + \ln (\pi) + 2 \ln(H/\mu) + 2 \ln(\Lambda_{\rm IR}/p) \right] 
\end{align}
with
\begin{align}
    \rho_{\rm logs}(\omega) = \rho_{\rm tree}(\omega)\, \frac{\omega^2}{4} \,, \quad \text{and} \quad \phi_{\rm logs}(\omega) = \phi_{\rm tree}(\omega)
\end{align}
as well as

\begin{align}
    q^2(\omega) = & \,\frac{\omega^2\left(\omega ^2+7\right) + 9/4 }{\omega^2 \left(\omega^2+9\right)} \,, \nonumber
         \\ \varphi(\omega) = &\, \arg\left[2\omega^3+2 i \omega^2+15\omega+9 i \right]\,.
\end{align}

\subsection{Sharp features}

The one-loop correction from sharp features is given by Eq.~\eqref{eq:result_integral_P_sharp_loop}, with
\begin{align}
    \frac{\Delta \mathcal{P}_{\rm sharp}^{\rm logs} (p)}{\mathcal{P}_0^2}  & =  A_{\rm sharp}  \Re\left\{\frac{i}{4 n^2(n-1) x} \left(1-2 i x/n\right)^{-3-n} q_n^{(4)}(x) \right\}  \nonumber \\
    & \quad \times \left[\gamma_E + \ln (\pi) + 2 \ln(H/\mu) + 2 \ln(\Lambda_{\rm IR}/p) \right] \,, \\
    \text{where} \quad q_n^{(4)}(x) &= -n^3 + 2 i n^2 (3 + n) x + 2 n (2 + n) (3 + n) x^2  \nonumber\\
  & \quad  + 4 i (1 + n) [-2 + (-3 + n) n] x^3 - 8 n^2 (1 + n) x^4  \,, \nonumber
\end{align}
as well as
\begin{align}
    p_n^{(5)}(x) = & \, -\frac{1}{n^4(n-1)} \left[ i n^5 + 2 n^4 (3 + n) x - 2 i n^3 \left[4 + n (3 + 5 n)\right] x^2 \right. \\ 
    & - 
 4 n^2 \left\{-4 + n \left[5 + (10 - 3 n) n\right]\right\} x^3 + 
 8 i n \left(-6 + n \left\{4 + n \left[4 + (n-1) n\right] \right\} \right) x^4 \nonumber \\
 & \left. + 32 (n-1) x^5 \right] \,. \nonumber
\end{align}
Moreover, the following simplification exists:
\begin{equation}
    z^{n}\Phi\left[z, 1, n\right] = - \ln \left(1-z\right) -\sum_{k=1}^{n-1} \frac{z^k}{k} \,, \quad \text{for} \quad z = \left(1-2 i x/n\right)^{-1} \,.
\end{equation}

Note that the logs' term can also be rewritten in the scaling limit $x\sim \sqrt{n}$, to give:
\begin{equation}
    \frac{\Delta \mathcal{P}_{\rm sharp}^{\rm logs} (p)}{\mathcal{P}_0^2 A_{\rm sharp}}  \underset{n\gg 1\,, \,  x^2/n \, \text{fixed}}{=} \Re\left[- 2 i x^3 e^{2 ix} \right] e^{-\frac{2 x^2}{n}} \times \left[\gamma_E + \ln (\pi) + 2 \ln(H/\mu) + 2 \ln(\Lambda_{\rm IR}/p) \right] \,,
\end{equation}
whose overall $x$-dependent factor is nothing but $2/3$ the one of the finite pieces.
\begin{figure}
        \centering        
        \includegraphics[width=\textwidth]{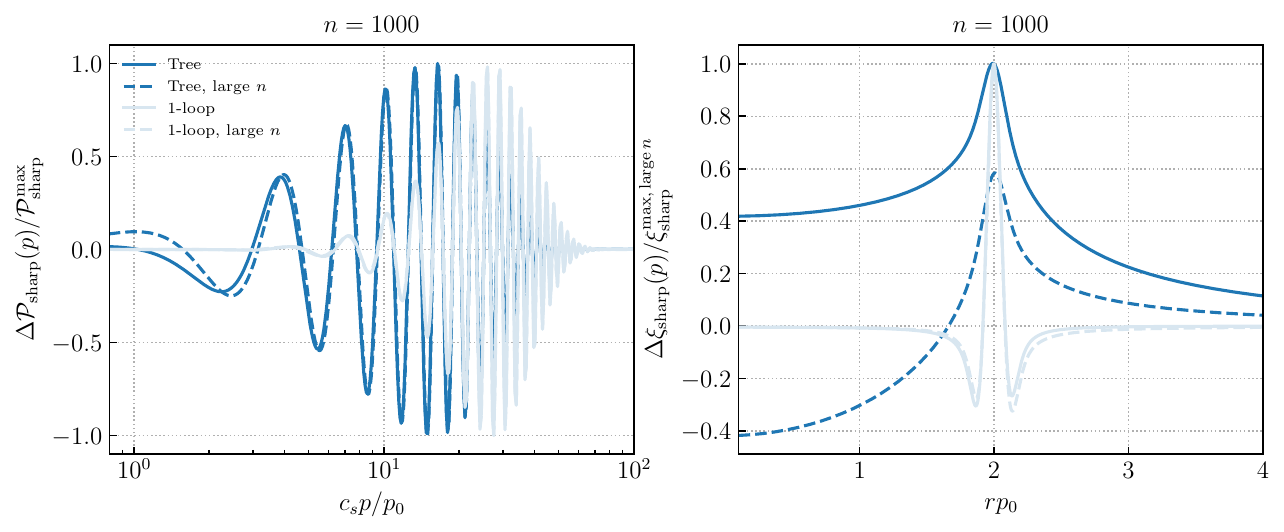}
        \caption{[Left] comparison of the tree-level and 1-loop corrections to the primordial power spectrum analytical results (solid lines) and their large $n$ (or equivalently small $\Delta N$ expansions. [Right] Same for the correction to the real-space 2-point correlation function. In the plots, we fix $n=1000$.  In both panels, the quantities are normalised to the maximum amplitudes calculated for the large $n$ limit, see Eqs.~\eqref{eq:P_max_tree},~\eqref{eq:P_max_loop} and~\eqref{eq:xi_max}. }
        \label{fig:large-n}
\end{figure}

Finally, in Fig.~\ref{fig:large-n}  we compare the large-$n$ expressions to the analytical ones. We can see that large $n$ expansion of the 1-loop correction to the primordial power spectrum--obtained neglecting the $\Phi$ functions--agrees extremely well with the analytical solution. The agreement translates into a very good matching of the corresponding  real-space 2pt  functions, where the one for the analytical result is calculated numerically.

On the other hand, for the tree-level results, the matching is very good only for $c_s p / p_0\ll1$. While this is  enough to estimate analytically the maximum amplitude of the correction to the power spectrum in the large $n$ limit, as it peaks at $c_s p^{\rm tree}_{\rm max} / p_0=\sqrt{n}/2$, the real-space 2pt functions, which receive contributions for small $c_s p / p_0\ll1$ are quite different. In particular, we find numerically that their maximum amplitudes are related by $ \Delta \xi_{\rm max}^{\rm tree,\,exact}\sim  \Delta \xi_{\rm max}^{{\rm tree,\,large }\,n}+1$, a difference which is irrelevant for $n\rightarrow \infty$ and in practice to set a perturbativity bound at very large $n$, but which is visible for the representative example $n=1000$ shown here.

\subsection{Time evolution of the 1-loop correction to the power spectrum}
\begin{figure}
        \centering
        \includegraphics[width=\textwidth]{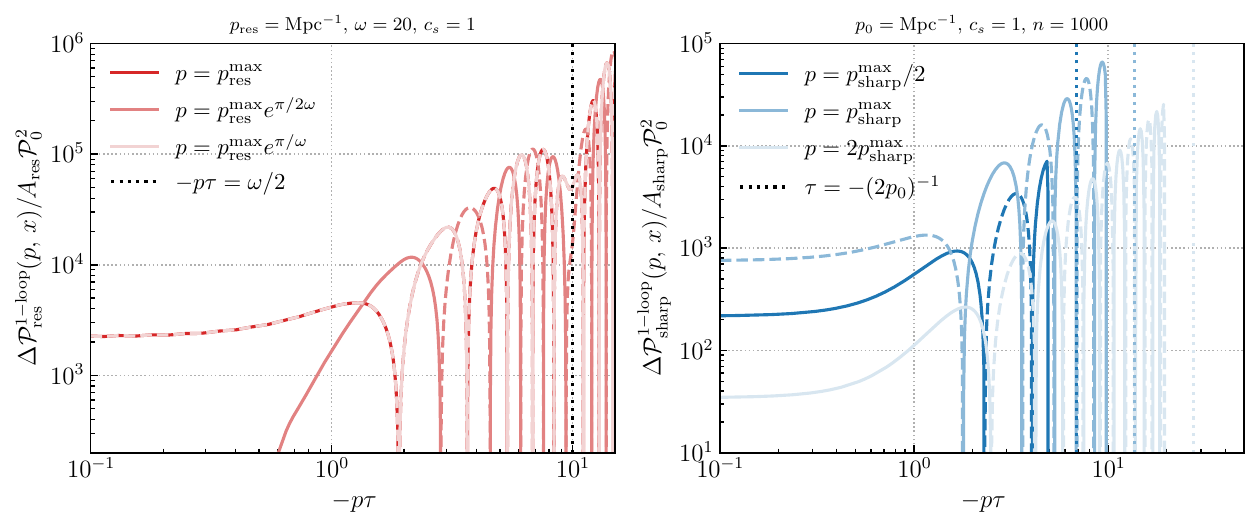}
        \caption{Time evolution of the 1-loop correction to the primordial power spectrum for resonant [left] and sharp [right] features. We plot 3 representative modes in each panel. In the left panel, we mark the resonance time $-p\tau=\omega/2$ with vertical lines. In the right panel, we also mark the time $\tau=-(2 p_0)^{-1}$ with vertical lines. We remind the reader that $p_{\rm res}^{\rm max}=p_{\rm res}/(2 c_s)\exp\left[-\phi_{\rm 1 \text{-} loop} (\omega)/\omega\right]$ and $p_{\rm sharp}^{\rm max} \underset{\Delta N \rightarrow 0}{=}p_0 \sqrt{3\pi/2}/(2 c_s \Delta N)$.}
        \label{fig:time-evo}
\end{figure}

We show the time evolution of the 1-loop correction to the power spectrum for a few Fourier modes for resonant and sharp features in Fig.~\ref{fig:time-evo}, focusing on one illustrative choice of parameters in each scenario. 

In the resonant feature case, the 3 modes are chosen so that one is the maximum of the logarithmic oscillations, one is out of phase with the maximum, and the third corresponds to the zero of the oscillation. As can be seen, the modes experience a resonance when  $-p\tau=\omega/2$. The correction then freezes to positive/negative values for the out of phase modes, while it decays to zero for the  third mode.

In the sharp feature case, the 3 modes are chosen so that one is located at the maximum amplitude of the oscillation envelope, and the other two are half and twice the mode. The correction is initially zero, and it is excited shortly after $\tau\sim-(2p_0)^{-1}$, after which the modes freeze to a constant.

\bibliographystyle{JHEP}
\bibliography{biblio}

\providecommand{\href}[2]{#2}\begingroup\raggedright\begin{thebibliography}{10}

\bibitem{Weinberg:2005vy}
S.~Weinberg, \emph{{Quantum contributions to cosmological correlations}}, \href{https://doi.org/10.1103/PhysRevD.72.043514}{\emph{Phys. Rev. D} {\bfseries 72} (2005) 043514} [\href{https://arxiv.org/abs/hep-th/0506236}{{\ttfamily hep-th/0506236}}].

\bibitem{Weinberg:2006ac}
S.~Weinberg, \emph{{Quantum contributions to cosmological correlations. II. Can these corrections become large?}}, \href{https://doi.org/10.1103/PhysRevD.74.023508}{\emph{Phys. Rev. D} {\bfseries 74} (2006) 023508} [\href{https://arxiv.org/abs/hep-th/0605244}{{\ttfamily hep-th/0605244}}].

\bibitem{Senatore:2012ya}
L.~Senatore and M.~Zaldarriaga, \emph{{The constancy of $\zeta$ in single-clock Inflation at all loops}}, \href{https://doi.org/10.1007/JHEP09(2013)148}{\emph{JHEP} {\bfseries 09} (2013) 148} [\href{https://arxiv.org/abs/1210.6048}{{\ttfamily 1210.6048}}].

\bibitem{Assassi:2012et}
V.~Assassi, D.~Baumann and D.~Green, \emph{{Symmetries and Loops in Inflation}}, \href{https://doi.org/10.1007/JHEP02(2013)151}{\emph{JHEP} {\bfseries 02} (2013) 151} [\href{https://arxiv.org/abs/1210.7792}{{\ttfamily 1210.7792}}].

\bibitem{Braglia:2025cee}
M.~Braglia and L.~Pinol, \emph{{One-loop renormalization of the effective field theory of inflationary fluctuations from gravitational interactions}}, \href{https://doi.org/10.1103/f6q6-5jxb}{\emph{Phys. Rev. D} {\bfseries 113} (2026) 063513} [\href{https://arxiv.org/abs/2504.07926}{{\ttfamily 2504.07926}}].

\bibitem{Braglia:2025qrb}
M.~Braglia and L.~Pinol, \emph{{Freezing of the renormalized one-loop primordial scalar power spectrum}}, \href{https://doi.org/10.1103/5jkv-mj3w}{\emph{Phys. Rev. D} {\bfseries 113} (2026) L061302} [\href{https://arxiv.org/abs/2504.13136}{{\ttfamily 2504.13136}}].

\bibitem{Creminelli:2006xe}
P.~Creminelli, M.~A. Luty, A.~Nicolis and L.~Senatore, \emph{{Starting the Universe: Stable Violation of the Null Energy Condition and Non-standard Cosmologies}}, \href{https://doi.org/10.1088/1126-6708/2006/12/080}{\emph{JHEP} {\bfseries 12} (2006) 080} [\href{https://arxiv.org/abs/hep-th/0606090}{{\ttfamily hep-th/0606090}}].

\bibitem{Cheung:2007st}
C.~Cheung, P.~Creminelli, A.~L. Fitzpatrick, J.~Kaplan and L.~Senatore, \emph{{The Effective Field Theory of Inflation}}, \href{https://doi.org/10.1088/1126-6708/2008/03/014}{\emph{JHEP} {\bfseries 03} (2008) 014} [\href{https://arxiv.org/abs/0709.0293}{{\ttfamily 0709.0293}}].

\bibitem{Ballesteros:2025nhz}
G.~Ballesteros, J.~G. Egea and F.~Riccardi, \emph{{Finite parts of inflationary loops II: A streamlined UV in-in algorithm and distinguishable signatures}}, \href{https://doi.org/10.1007/JHEP07(2026)059}{\emph{JHEP} {\bfseries 07} (2026) 059} [\href{https://arxiv.org/abs/2512.20467}{{\ttfamily 2512.20467}}].

\bibitem{Senatore:2009cf}
L.~Senatore and M.~Zaldarriaga, \emph{{On Loops in Inflation}}, \href{https://doi.org/10.1007/JHEP12(2010)008}{\emph{JHEP} {\bfseries 12} (2010) 008} [\href{https://arxiv.org/abs/0912.2734}{{\ttfamily 0912.2734}}].

\bibitem{Marolf:2010zp}
D.~Marolf and I.~A. Morrison, \emph{{The IR stability of de Sitter: Loop corrections to scalar propagators}}, \href{https://doi.org/10.1103/PhysRevD.82.105032}{\emph{Phys. Rev. D} {\bfseries 82} (2010) 105032} [\href{https://arxiv.org/abs/1006.0035}{{\ttfamily 1006.0035}}].

\bibitem{Senatore:2012nq}
L.~Senatore and M.~Zaldarriaga, \emph{{On Loops in Inflation II: IR Effects in Single Clock Inflation}}, \href{https://doi.org/10.1007/JHEP01(2013)109}{\emph{JHEP} {\bfseries 01} (2013) 109} [\href{https://arxiv.org/abs/1203.6354}{{\ttfamily 1203.6354}}].

\bibitem{Premkumar:2021mlz}
A.~Premkumar, \emph{{Regulating loops in de Sitter spacetime}}, \href{https://doi.org/10.1103/PhysRevD.109.045003}{\emph{Phys. Rev. D} {\bfseries 109} (2024) 045003} [\href{https://arxiv.org/abs/2110.12504}{{\ttfamily 2110.12504}}].

\bibitem{Xianyu:2022jwk}
Z.-Z. Xianyu and H.~Zhang, \emph{{Bootstrapping one-loop inflation correlators with the spectral decomposition}}, \href{https://doi.org/10.1007/JHEP04(2023)103}{\emph{JHEP} {\bfseries 04} (2023) 103} [\href{https://arxiv.org/abs/2211.03810}{{\ttfamily 2211.03810}}].

\bibitem{Qin:2023bjk}
Z.~Qin and Z.-Z. Xianyu, \emph{{Inflation correlators at the one-loop order: nonanalyticity, factorization, cutting rule, and OPE}}, \href{https://doi.org/10.1007/JHEP09(2023)116}{\emph{JHEP} {\bfseries 09} (2023) 116} [\href{https://arxiv.org/abs/2304.13295}{{\ttfamily 2304.13295}}].

\bibitem{Qin:2024gtr}
Z.~Qin, \emph{{Cosmological correlators at the loop level}}, \href{https://doi.org/10.1007/JHEP03(2025)051}{\emph{JHEP} {\bfseries 03} (2025) 051} [\href{https://arxiv.org/abs/2411.13636}{{\ttfamily 2411.13636}}].

\bibitem{Cespedes:2025dnq}
S.~Cespedes and S.~Jazayeri, \emph{{The massive flat space limit of cosmological correlators}}, \href{https://doi.org/10.1007/JHEP07(2025)032}{\emph{JHEP} {\bfseries 07} (2025) 032} [\href{https://arxiv.org/abs/2501.02119}{{\ttfamily 2501.02119}}].

\bibitem{Cespedes:2025ple}
S.~Cespedes, Z.~Qin and D.-G. Wang, \emph{{{\ensuremath{\lambda}}{\ensuremath{\phi}}$^{4}$ as an effective theory in de Sitter}}, \href{https://doi.org/10.1007/JHEP05(2026)143}{\emph{JHEP} {\bfseries 05} (2026) 143} [\href{https://arxiv.org/abs/2510.25826}{{\ttfamily 2510.25826}}].

\bibitem{Pimentel:2026kqc}
G.~L. Pimentel and T.~Westerdijk, \emph{{On Cosmological Correlators at One Loop}},  \href{https://arxiv.org/abs/2601.00952}{{\ttfamily 2601.00952}}.

\bibitem{Ema:2026dop}
Y.~Ema, M.~Hong, R.~Jinno and K.~Mukaida, \emph{{Cancellation of loop corrections to soft scalar power spectrum}},  \href{https://arxiv.org/abs/2603.01961}{{\ttfamily 2603.01961}}.

\bibitem{Farren:2026hao}
A.~Farren, C.~McCulloch, E.~Pajer and X.~Tong, \emph{{All-Loop Renormalization and the Phase of the de Sitter Wavefunction}},  \href{https://arxiv.org/abs/2603.08794}{{\ttfamily 2603.08794}}.

\bibitem{Ballesteros:2024cef}
G.~Ballesteros, J.~Gamb{\'\i}n~Egea and F.~Riccardi, \emph{{Finite parts of inflationary loops}}, \href{https://doi.org/10.1007/JHEP06(2025)098}{\emph{JHEP} {\bfseries 06} (2025) 098} [\href{https://arxiv.org/abs/2411.19674}{{\ttfamily 2411.19674}}].

\bibitem{Chen:2010xka}
X.~Chen, \emph{{Primordial Non-Gaussianities from Inflation Models}}, \href{https://doi.org/10.1155/2010/638979}{\emph{Adv. Astron.} {\bfseries 2010} (2010) 638979} [\href{https://arxiv.org/abs/1002.1416}{{\ttfamily 1002.1416}}].

\bibitem{Achucarro:2022qrl}
A.~Ach{\'u}carro et~al., \emph{{Inflation: Theory and Observations}},  \href{https://arxiv.org/abs/2203.08128}{{\ttfamily 2203.08128}}.

\bibitem{Planck:2018jri}
{\scshape Planck} collaboration, \emph{{Planck 2018 results. X. Constraints on inflation}}, \href{https://doi.org/10.1051/0004-6361/201833887}{\emph{Astron. Astrophys.} {\bfseries 641} (2020) A10} [\href{https://arxiv.org/abs/1807.06211}{{\ttfamily 1807.06211}}].

\bibitem{Chen:2014cwa}
X.~Chen, M.~H. Namjoo and Y.~Wang, \emph{{Models of the Primordial Standard Clock}}, \href{https://doi.org/10.1088/1475-7516/2015/02/027}{\emph{JCAP} {\bfseries 02} (2015) 027} [\href{https://arxiv.org/abs/1411.2349}{{\ttfamily 1411.2349}}].

\bibitem{Braglia:2021ckn}
M.~Braglia, X.~Chen and D.~K. Hazra, \emph{{Comparing multi-field primordial feature models with the Planck data}}, \href{https://doi.org/10.1088/1475-7516/2021/06/005}{\emph{JCAP} {\bfseries 06} (2021) 005} [\href{https://arxiv.org/abs/2103.03025}{{\ttfamily 2103.03025}}].

\bibitem{Pinol:2024arz}
L.~Pinol, \emph{{Effective field theory of multifield inflationary fluctuations}}, \href{https://doi.org/10.1103/PhysRevD.110.L041302}{\emph{Phys. Rev. D} {\bfseries 110} (2024) L041302} [\href{https://arxiv.org/abs/2405.02190}{{\ttfamily 2405.02190}}].

\bibitem{Ballesteros:2021fsp}
G.~Ballesteros, S.~C{\'e}spedes and L.~Santoni, \emph{{Large power spectrum and primordial black holes in the effective theory of inflation}}, \href{https://doi.org/10.1007/JHEP01(2022)074}{\emph{JHEP} {\bfseries 01} (2022) 074} [\href{https://arxiv.org/abs/2109.00567}{{\ttfamily 2109.00567}}].

\bibitem{CarrilloGonzalez:2025fqq}
M.~Carrillo~Gonz{\'a}lez and S.~C{\'e}spedes, \emph{{Causality bounds on the primordial power spectrum}}, \href{https://doi.org/10.1088/1475-7516/2025/08/071}{\emph{JCAP} {\bfseries 08} (2025) 071} [\href{https://arxiv.org/abs/2502.19477}{{\ttfamily 2502.19477}}].

\bibitem{Braglia:2024zsl}
M.~Braglia and L.~Pinol, \emph{{No time to derive: unraveling total time derivatives in in-in perturbation theory}}, \href{https://doi.org/10.1007/JHEP08(2024)068}{\emph{JHEP} {\bfseries 08} (2024) 068} [\href{https://arxiv.org/abs/2403.14558}{{\ttfamily 2403.14558}}].

\bibitem{Urakawa:2010it}
Y.~Urakawa and T.~Tanaka, \emph{{IR divergence does not affect the gauge-invariant curvature perturbation}}, \href{https://doi.org/10.1103/PhysRevD.82.121301}{\emph{Phys. Rev. D} {\bfseries 82} (2010) 121301} [\href{https://arxiv.org/abs/1007.0468}{{\ttfamily 1007.0468}}].

\bibitem{Urakawa:2010kr}
Y.~Urakawa and T.~Tanaka, \emph{{Natural selection of inflationary vacuum required by infra-red regularity and gauge-invariance}}, \href{https://doi.org/10.1143/PTP.125.1067}{\emph{Prog. Theor. Phys.} {\bfseries 125} (2011) 1067} [\href{https://arxiv.org/abs/1009.2947}{{\ttfamily 1009.2947}}].

\bibitem{Giddings:2010nc}
S.~B. Giddings and M.~S. Sloth, \emph{{Semiclassical relations and IR effects in de Sitter and slow-roll space-times}}, \href{https://doi.org/10.1088/1475-7516/2011/01/023}{\emph{JCAP} {\bfseries 01} (2011) 023} [\href{https://arxiv.org/abs/1005.1056}{{\ttfamily 1005.1056}}].

\bibitem{Gerstenlauer:2011ti}
M.~Gerstenlauer, A.~Hebecker and G.~Tasinato, \emph{{Inflationary Correlation Functions without Infrared Divergences}}, \href{https://doi.org/10.1088/1475-7516/2011/06/021}{\emph{JCAP} {\bfseries 06} (2011) 021} [\href{https://arxiv.org/abs/1102.0560}{{\ttfamily 1102.0560}}].

\bibitem{Tanaka:2011aj}
T.~Tanaka and Y.~Urakawa, \emph{{Dominance of gauge artifact in the consistency relation for the primordial bispectrum}}, \href{https://doi.org/10.1088/1475-7516/2011/05/014}{\emph{JCAP} {\bfseries 05} (2011) 014} [\href{https://arxiv.org/abs/1103.1251}{{\ttfamily 1103.1251}}].

\bibitem{Pajer:2013ana}
E.~Pajer, F.~Schmidt and M.~Zaldarriaga, \emph{{The Observed Squeezed Limit of Cosmological Three-Point Functions}}, \href{https://doi.org/10.1103/PhysRevD.88.083502}{\emph{Phys. Rev. D} {\bfseries 88} (2013) 083502} [\href{https://arxiv.org/abs/1305.0824}{{\ttfamily 1305.0824}}].

\bibitem{Seery:2010kh}
D.~Seery, \emph{{Infrared effects in inflationary correlation functions}}, \href{https://doi.org/10.1088/0264-9381/27/12/124005}{\emph{Class. Quant. Grav.} {\bfseries 27} (2010) 124005} [\href{https://arxiv.org/abs/1005.1649}{{\ttfamily 1005.1649}}].

\bibitem{Noumi:2012vr}
T.~Noumi, M.~Yamaguchi and D.~Yokoyama, \emph{{Effective field theory approach to quasi-single field inflation and effects of heavy fields}}, \href{https://doi.org/10.1007/JHEP06(2013)051}{\emph{JHEP} {\bfseries 06} (2013) 051} [\href{https://arxiv.org/abs/1211.1624}{{\ttfamily 1211.1624}}].

\bibitem{Pimentel:2012tw}
G.~L. Pimentel, L.~Senatore and M.~Zaldarriaga, \emph{{On Loops in Inflation III: Time Independence of zeta in Single Clock Inflation}}, \href{https://doi.org/10.1007/JHEP07(2012)166}{\emph{JHEP} {\bfseries 07} (2012) 166} [\href{https://arxiv.org/abs/1203.6651}{{\ttfamily 1203.6651}}].

\bibitem{Braglia:2022ftm}
M.~Braglia, X.~Chen, D.~K. Hazra and L.~Pinol, \emph{{Back to the features: assessing the discriminating power of future CMB missions on inflationary models}}, \href{https://doi.org/10.1088/1475-7516/2023/03/014}{\emph{JCAP} {\bfseries 03} (2023) 014} [\href{https://arxiv.org/abs/2210.07028}{{\ttfamily 2210.07028}}].

\bibitem{Chen:2008wn}
X.~Chen, R.~Easther and E.~A. Lim, \emph{{Generation and Characterization of Large Non-Gaussianities in Single Field Inflation}}, \href{https://doi.org/10.1088/1475-7516/2008/04/010}{\emph{JCAP} {\bfseries 04} (2008) 010} [\href{https://arxiv.org/abs/0801.3295}{{\ttfamily 0801.3295}}].

\bibitem{Flauger:2009ab}
R.~Flauger, L.~McAllister, E.~Pajer, A.~Westphal and G.~Xu, \emph{{Oscillations in the CMB from Axion Monodromy Inflation}}, \href{https://doi.org/10.1088/1475-7516/2010/06/009}{\emph{JCAP} {\bfseries 06} (2010) 009} [\href{https://arxiv.org/abs/0907.2916}{{\ttfamily 0907.2916}}].

\bibitem{Flauger:2010ja}
R.~Flauger and E.~Pajer, \emph{{Resonant Non-Gaussianity}}, \href{https://doi.org/10.1088/1475-7516/2011/01/017}{\emph{JCAP} {\bfseries 01} (2011) 017} [\href{https://arxiv.org/abs/1002.0833}{{\ttfamily 1002.0833}}].

\bibitem{McAllister:2008hb}
L.~McAllister, E.~Silverstein and A.~Westphal, \emph{{Gravity Waves and Linear Inflation from Axion Monodromy}}, \href{https://doi.org/10.1103/PhysRevD.82.046003}{\emph{Phys. Rev. D} {\bfseries 82} (2010) 046003} [\href{https://arxiv.org/abs/0808.0706}{{\ttfamily 0808.0706}}].

\bibitem{Chen:2010bka}
X.~Chen, \emph{{Folded Resonant Non-Gaussianity in General Single Field Inflation}}, \href{https://doi.org/10.1088/1475-7516/2010/12/003}{\emph{JCAP} {\bfseries 12} (2010) 003} [\href{https://arxiv.org/abs/1008.2485}{{\ttfamily 1008.2485}}].

\bibitem{Behbahani:2011it}
S.~R. Behbahani, A.~Dymarsky, M.~Mirbabayi and L.~Senatore, \emph{{(Small) Resonant non-Gaussianities: Signatures of a Discrete Shift Symmetry in the Effective Field Theory of Inflation}}, \href{https://doi.org/10.1088/1475-7516/2012/12/036}{\emph{JCAP} {\bfseries 12} (2012) 036} [\href{https://arxiv.org/abs/1111.3373}{{\ttfamily 1111.3373}}].

\bibitem{Chen:2015dga}
X.~Chen, M.~H. Namjoo and Y.~Wang, \emph{{On the equation-of-motion versus in-in approach in cosmological perturbation theory}}, \href{https://doi.org/10.1088/1475-7516/2016/01/022}{\emph{JCAP} {\bfseries 01} (2016) 022} [\href{https://arxiv.org/abs/1505.03955}{{\ttfamily 1505.03955}}].

\bibitem{Beneke:1997zp}
M.~Beneke and V.~A. Smirnov, \emph{{Asymptotic expansion of Feynman integrals near threshold}}, \href{https://doi.org/10.1016/S0550-3213(98)00138-2}{\emph{Nucl. Phys. B} {\bfseries 522} (1998) 321} [\href{https://arxiv.org/abs/hep-ph/9711391}{{\ttfamily hep-ph/9711391}}].

\bibitem{Smirnov:1999bza}
V.~A. Smirnov, \emph{{Problems of the strategy of regions}}, \href{https://doi.org/10.1016/S0370-2693(99)01061-8}{\emph{Phys. Lett. B} {\bfseries 465} (1999) 226} [\href{https://arxiv.org/abs/hep-ph/9907471}{{\ttfamily hep-ph/9907471}}].

\bibitem{Beneke:2023wmt}
M.~Beneke, P.~Hager and A.~F. Sanfilippo, \emph{{Cosmological correlators in massless {\ensuremath{\phi}}$^{4}$-theory and the method of regions}}, \href{https://doi.org/10.1007/JHEP04(2024)006}{\emph{JHEP} {\bfseries 04} (2024) 006} [\href{https://arxiv.org/abs/2312.06766}{{\ttfamily 2312.06766}}].

\bibitem{Beneke:2026rtf}
M.~Beneke, P.~Hager and A.~F. Sanfilippo, \emph{{Renormalisation and matching of massless scalar correlation functions in Soft de Sitter Effective Theory}}, \href{https://doi.org/10.1007/JHEP07(2026)177}{\emph{JHEP} {\bfseries 07} (2026) 177} [\href{https://arxiv.org/abs/2603.09438}{{\ttfamily 2603.09438}}].

\bibitem{Beneke:2026ksj}
M.~Beneke, P.~Hager and A.~F. Sanfilippo, \emph{{Quantum correction to the diffusion term in stochastic inflation from composite-operator matching in Soft de Sitter Effective Theory}},  \href{https://arxiv.org/abs/2604.14283}{{\ttfamily 2604.14283}}.

\bibitem{Cespedes:2012hu}
S.~Cespedes, V.~Atal and G.~A. Palma, \emph{{On the importance of heavy fields during inflation}}, \href{https://doi.org/10.1088/1475-7516/2012/05/008}{\emph{JCAP} {\bfseries 05} (2012) 008} [\href{https://arxiv.org/abs/1201.4848}{{\ttfamily 1201.4848}}].

\bibitem{Bartolo:2013exa}
N.~Bartolo, D.~Cannone and S.~Matarrese, \emph{{The Effective Field Theory of Inflation Models with Sharp Features}}, \href{https://doi.org/10.1088/1475-7516/2013/10/038}{\emph{JCAP} {\bfseries 10} (2013) 038} [\href{https://arxiv.org/abs/1307.3483}{{\ttfamily 1307.3483}}].

\bibitem{Cannone:2014qna}
D.~Cannone, N.~Bartolo and S.~Matarrese, \emph{{Perturbative Unitarity of Inflationary Models with Features}}, \href{https://doi.org/10.1103/PhysRevD.89.127301}{\emph{Phys. Rev. D} {\bfseries 89} (2014) 127301} [\href{https://arxiv.org/abs/1402.2258}{{\ttfamily 1402.2258}}].

\bibitem{Cheng:2021lif}
S.-L. Cheng, D.-S. Lee and K.-W. Ng, \emph{{Power spectrum of primordial perturbations during ultra-slow-roll inflation with back reaction effects}}, \href{https://doi.org/10.1016/j.physletb.2022.136956}{\emph{Phys. Lett. B} {\bfseries 827} (2022) 136956} [\href{https://arxiv.org/abs/2106.09275}{{\ttfamily 2106.09275}}].

\bibitem{Inomata:2022yte}
K.~Inomata, M.~Braglia, X.~Chen and S.~Renaux-Petel, \emph{{Questions on calculation of primordial power spectrum with large spikes: the resonance model case}}, \href{https://doi.org/10.1088/1475-7516/2023/04/011}{\emph{JCAP} {\bfseries 04} (2023) 011} [\href{https://arxiv.org/abs/2211.02586}{{\ttfamily 2211.02586}}].

\bibitem{Kristiano:2022maq}
J.~Kristiano and J.~Yokoyama, \emph{{Constraining Primordial Black Hole Formation from Single-Field Inflation}}, \href{https://doi.org/10.1103/PhysRevLett.132.221003}{\emph{Phys. Rev. Lett.} {\bfseries 132} (2024) 221003} [\href{https://arxiv.org/abs/2211.03395}{{\ttfamily 2211.03395}}].

\bibitem{Riotto:2023hoz}
A.~Riotto, \emph{{The Primordial Black Hole Formation from Single-Field Inflation is Not Ruled Out}},  \href{https://arxiv.org/abs/2301.00599}{{\ttfamily 2301.00599}}.

\bibitem{Firouzjahi:2023aum}
H.~Firouzjahi, \emph{{One-loop corrections in power spectrum in single field inflation}}, \href{https://doi.org/10.1088/1475-7516/2023/10/006}{\emph{JCAP} {\bfseries 10} (2023) 006} [\href{https://arxiv.org/abs/2303.12025}{{\ttfamily 2303.12025}}].

\bibitem{Choudhury:2023vuj}
S.~Choudhury, M.~R. Gangopadhyay and M.~Sami, \emph{{No-go for the formation of heavy mass Primordial Black Holes in Single Field Inflation}}, \href{https://doi.org/10.1140/epjc/s10052-024-13218-2}{\emph{Eur. Phys. J. C} {\bfseries 84} (2024) 884} [\href{https://arxiv.org/abs/2301.10000}{{\ttfamily 2301.10000}}].

\bibitem{Motohashi:2023syh}
H.~Motohashi and Y.~Tada, \emph{{Squeezed bispectrum and one-loop corrections in transient constant-roll inflation}}, \href{https://doi.org/10.1088/1475-7516/2023/08/069}{\emph{JCAP} {\bfseries 08} (2023) 069} [\href{https://arxiv.org/abs/2303.16035}{{\ttfamily 2303.16035}}].

\bibitem{Franciolini:2023lgy}
G.~Franciolini, A.~Iovino, Junior., M.~Taoso and A.~Urbano, \emph{{Perturbativity in the presence of ultraslow-roll dynamics}}, \href{https://doi.org/10.1103/PhysRevD.109.123550}{\emph{Phys. Rev. D} {\bfseries 109} (2024) 123550} [\href{https://arxiv.org/abs/2305.03491}{{\ttfamily 2305.03491}}].

\bibitem{Tasinato:2023ukp}
G.~Tasinato, \emph{{Large |{\ensuremath{\eta}}| approach to single field inflation}}, \href{https://doi.org/10.1103/PhysRevD.108.043526}{\emph{Phys. Rev. D} {\bfseries 108} (2023) 043526} [\href{https://arxiv.org/abs/2305.11568}{{\ttfamily 2305.11568}}].

\bibitem{Cheng:2023ikq}
S.-L. Cheng, D.-S. Lee and K.-W. Ng, \emph{{Primordial perturbations from ultra-slow-roll single-field inflation with quantum loop effects}}, \href{https://doi.org/10.1088/1475-7516/2024/03/008}{\emph{JCAP} {\bfseries 03} (2024) 008} [\href{https://arxiv.org/abs/2305.16810}{{\ttfamily 2305.16810}}].

\bibitem{Fumagalli:2023hpa}
J.~Fumagalli, \emph{{Absence of one-loop effects on large scales from small scales in non-slow-roll dynamics}}, \href{https://doi.org/10.1007/JHEP05(2025)162}{\emph{JHEP} {\bfseries 05} (2025) 162} [\href{https://arxiv.org/abs/2305.19263}{{\ttfamily 2305.19263}}].

\bibitem{Maity:2023qzw}
S.~Maity, H.~V. Ragavendra, S.~K. Sethi and L.~Sriramkumar, \emph{{Loop contributions to the scalar power spectrum due to quartic order action in ultra slow roll inflation}}, \href{https://doi.org/10.1088/1475-7516/2024/05/046}{\emph{JCAP} {\bfseries 05} (2024) 046} [\href{https://arxiv.org/abs/2307.13636}{{\ttfamily 2307.13636}}].

\bibitem{Davies:2023hhn}
M.~W. Davies, L.~Iacconi and D.~J. Mulryne, \emph{{Numerical 1-loop correction from a potential yielding ultra-slow-roll dynamics}}, \href{https://doi.org/10.1088/1475-7516/2024/04/050}{\emph{JCAP} {\bfseries 04} (2024) 050} [\href{https://arxiv.org/abs/2312.05694}{{\ttfamily 2312.05694}}].

\bibitem{Tada:2023rgp}
Y.~Tada, T.~Terada and J.~Tokuda, \emph{{Cancellation of quantum corrections on the soft curvature perturbations}}, \href{https://doi.org/10.1007/JHEP01(2024)105}{\emph{JHEP} {\bfseries 01} (2024) 105} [\href{https://arxiv.org/abs/2308.04732}{{\ttfamily 2308.04732}}].

\bibitem{Iacconi:2023ggt}
L.~Iacconi, D.~Mulryne and D.~Seery, \emph{{Loop corrections in the separate universe picture}}, \href{https://doi.org/10.1088/1475-7516/2024/06/062}{\emph{JCAP} {\bfseries 06} (2024) 062} [\href{https://arxiv.org/abs/2312.12424}{{\ttfamily 2312.12424}}].

\bibitem{Inomata:2024lud}
K.~Inomata, \emph{{Superhorizon Curvature Perturbations Are Protected against One-Loop Corrections}}, \href{https://doi.org/10.1103/PhysRevLett.133.141001}{\emph{Phys. Rev. Lett.} {\bfseries 133} (2024) 141001} [\href{https://arxiv.org/abs/2403.04682}{{\ttfamily 2403.04682}}].

\bibitem{Firouzjahi:2024psd}
H.~Firouzjahi, \emph{{Loop corrections in the bispectrum in ultraslow-roll inflation with PBHs formation}}, \href{https://doi.org/10.1103/PhysRevD.110.043519}{\emph{Phys. Rev. D} {\bfseries 110} (2024) 043519} [\href{https://arxiv.org/abs/2403.03841}{{\ttfamily 2403.03841}}].

\bibitem{Caravano:2024tlp}
A.~Caravano, K.~Inomata and S.~Renaux-Petel, \emph{{Inflationary Butterfly Effect: Nonperturbative Dynamics from Small-Scale Features}}, \href{https://doi.org/10.1103/PhysRevLett.133.151001}{\emph{Phys. Rev. Lett.} {\bfseries 133} (2024) 151001} [\href{https://arxiv.org/abs/2403.12811}{{\ttfamily 2403.12811}}].

\bibitem{Ballesteros:2024zdp}
G.~Ballesteros and J.~Gamb{\'\i}n~Egea, \emph{{One-loop power spectrum in ultra slow-roll inflation and implications for primordial black hole dark matter}}, \href{https://doi.org/10.1088/1475-7516/2024/07/052}{\emph{JCAP} {\bfseries 07} (2024) 052} [\href{https://arxiv.org/abs/2404.07196}{{\ttfamily 2404.07196}}].

\bibitem{Kristiano:2024vst}
J.~Kristiano and J.~Yokoyama, \emph{{Comparing sharp and smooth transitions of the second slow-roll parameter in single-field inflation}}, \href{https://doi.org/10.1088/1475-7516/2024/10/036}{\emph{JCAP} {\bfseries 10} (2024) 036} [\href{https://arxiv.org/abs/2405.12145}{{\ttfamily 2405.12145}}].

\bibitem{Kawaguchi:2024rsv}
R.~Kawaguchi, S.~Tsujikawa and Y.~Yamada, \emph{{Proving the absence of large one-loop corrections to the power spectrum of curvature perturbations in transient ultra-slow-roll inflation within the path-integral approach}}, \href{https://doi.org/10.1007/JHEP12(2024)095}{\emph{JHEP} {\bfseries 12} (2024) 095} [\href{https://arxiv.org/abs/2407.19742}{{\ttfamily 2407.19742}}].

\bibitem{Fumagalli:2024jzz}
J.~Fumagalli, \emph{{Absence of one-loop effects on large scales from small scales in non-slow-roll dynamics. Part 2. Quartic interactions and consistency relations}}, \href{https://doi.org/10.1007/JHEP01(2025)108}{\emph{JHEP} {\bfseries 01} (2025) 108} [\href{https://arxiv.org/abs/2408.08296}{{\ttfamily 2408.08296}}].

\bibitem{Caravano:2024moy}
A.~Caravano, G.~Franciolini and S.~Renaux-Petel, \emph{{Ultraslow-roll inflation on the lattice: Backreaction and nonlinear effects}}, \href{https://doi.org/10.1103/PhysRevD.111.063518}{\emph{Phys. Rev. D} {\bfseries 111} (2025) 063518} [\href{https://arxiv.org/abs/2410.23942}{{\ttfamily 2410.23942}}].

\bibitem{Ruiz:2024weh}
J.~{\'A}. Ruiz and J.~Rey, \emph{{Gravitational waves in ultra-slow-roll and their anisotropy at two loops}}, \href{https://doi.org/10.1088/1475-7516/2025/04/026}{\emph{JCAP} {\bfseries 04} (2025) 026} [\href{https://arxiv.org/abs/2410.09014}{{\ttfamily 2410.09014}}].

\bibitem{Firouzjahi:2024sce}
H.~Firouzjahi, \emph{{Two-Loop Corrections in Power Spectrum in Models of Inflation with Primordial Black Hole Formation}}, \href{https://doi.org/10.3390/universe10120456}{\emph{Universe} {\bfseries 10} (2024) 456} [\href{https://arxiv.org/abs/2411.10253}{{\ttfamily 2411.10253}}].

\bibitem{Sheikhahmadi:2024peu}
H.~Sheikhahmadi and A.~Nassiri-Rad, \emph{{RG-Flow Renormalized One-Loop Corrections to the Power Spectrum in USR Inflation}},  \href{https://arxiv.org/abs/2411.18525}{{\ttfamily 2411.18525}}.

\bibitem{Inomata:2025bqw}
K.~Inomata, \emph{{Conservation of superhorizon curvature perturbations at one loop: Backreaction in the in-in formalism and renormalization}}, \href{https://doi.org/10.1103/PhysRevD.111.103504}{\emph{Phys. Rev. D} {\bfseries 111} (2025) 103504} [\href{https://arxiv.org/abs/2502.08707}{{\ttfamily 2502.08707}}].

\bibitem{Fang:2025vhi}
C.-J. Fang, Z.-H. Lyu, C.~Chen and Z.-K. Guo, \emph{{Incorporating backreaction in one-loop corrections in ultraslow-roll inflation}}, \href{https://doi.org/10.1103/nkrq-3d39}{\emph{Phys. Rev. D} {\bfseries 112} (2025) 023547} [\href{https://arxiv.org/abs/2502.09555}{{\ttfamily 2502.09555}}].

\bibitem{Firouzjahi:2025gja}
H.~Firouzjahi and B.~Nikbakht, \emph{{Nonperturbative Hamiltonian and higher loop corrections in ultraslow-roll inflation}}, \href{https://doi.org/10.1103/npwq-dz79}{\emph{Phys. Rev. D} {\bfseries 113} (2026) L061301} [\href{https://arxiv.org/abs/2502.09481}{{\ttfamily 2502.09481}}].

\bibitem{Firouzjahi:2025ihn}
H.~Firouzjahi and B.~Nikbakht, \emph{{Hamiltonians to all orders in perturbation theory and higher-loop corrections in single-field inflation with PBH formation}}, \href{https://doi.org/10.1103/zt8h-862w}{\emph{Phys. Rev. D} {\bfseries 113} (2026) 063510} [\href{https://arxiv.org/abs/2502.10287}{{\ttfamily 2502.10287}}].

\bibitem{Inomata:2025pqa}
K.~Inomata, \emph{{Role of the counterterms in the conservation of superhorizon curvature perturbations at one loop}}, \href{https://doi.org/10.1103/r8bh-s48f}{\emph{Phys. Rev. D} {\bfseries 111} (2025) 123517} [\href{https://arxiv.org/abs/2502.12112}{{\ttfamily 2502.12112}}].

\bibitem{Iacconi:2026uzo}
L.~Iacconi, D.~Mulryne and D.~Seery, \emph{{Decoupling of large-scale, adiabatic inflationary perturbations from enhanced small-scale modes at one-loop}},  \href{https://arxiv.org/abs/2601.14229}{{\ttfamily 2601.14229}}.

\bibitem{Li:2026vrn}
W.~Li and C.~Chen, \emph{{Evolution of Linear Perturbations under Time-Dependent Hubble Friction I: SR-USR-SR Inflation}},  \href{https://arxiv.org/abs/2602.13074}{{\ttfamily 2602.13074}}.

\bibitem{Bhowmick:2024kld}
S.~Bhowmick, D.~Ghosh and F.~Ullah, \emph{{Bispectrum at 1-loop in the Effective Field Theory of Inflation}}, \href{https://doi.org/10.1007/JHEP10(2024)057}{\emph{JHEP} {\bfseries 10} (2024) 057} [\href{https://arxiv.org/abs/2405.10374}{{\ttfamily 2405.10374}}].

\bibitem{ParkerFulling1974}
L.~Parker and S.~A. Fulling, \emph{Adiabatic regularization of the energy-momentum tensor of a quantized field in homogeneous spaces}, \href{https://doi.org/10.1103/PhysRevD.9.341}{\emph{Phys. Rev. D} {\bfseries 9} (1974) 341}.

\bibitem{BirrellDavies}
N.~D. Birrell and P.~C.~W. Davies, \emph{Quantum Fields in Curved Space}. Cambridge University Press, 1982.

\bibitem{AndersonParker1987}
P.~R. Anderson and L.~Parker, \emph{Adiabatic regularization in closed robertson-walker universes}, \href{https://doi.org/10.1103/PhysRevD.36.2963}{\emph{Phys. Rev. D} {\bfseries 36} (1987) 2963}.

\bibitem{Kristiano:2025ajj}
J.~Kristiano and J.~Yokoyama, \emph{{Inflationary background renormalization}},  \href{https://arxiv.org/abs/2504.18514}{{\ttfamily 2504.18514}}.

\end{thebibliography}\endgroup

\end{document}